\pdfoutput=1
\documentclass[11pt]{article}
\usepackage[a4paper,margin=2.3cm]{geometry}
\usepackage{mathptmx}
\usepackage{amsmath}
\usepackage{newtxmath}   %
\usepackage{bm}          %
\usepackage{graphicx}
\usepackage{booktabs,tabularx,array}
\usepackage{xcolor}
\usepackage[font=small,labelfont=bf,labelsep=period,justification=justified,singlelinecheck=false]{caption}
\usepackage{setspace}
\usepackage{authblk}
\usepackage[super,sort&compress,comma]{natbib}
\usepackage[colorlinks=true,linkcolor=blue!50!black,citecolor=blue!50!black,urlcolor=blue!50!black]{hyperref}
\newcommand{\doi}[1]{doi: \begingroup\urlstyle{rm}\href{https://doi.org/#1}{\nolinkurl{#1}}\endgroup}
\usepackage{titlesec}
\usepackage{placeins}
\titleformat{\section}{\large\bfseries}{}{0pt}{}
\titleformat{\subsection}{\normalsize\bfseries}{}{0pt}{}
\titlespacing*{\section}{0pt}{1.2em}{0.5em}
\titlespacing*{\subsection}{0pt}{0.9em}{0.3em}
\graphicspath{{figures/}}

\newif\ifdraft
\draftfalse

\newcommand{\hb}{{\mathchar'26\mkern-9mu h}}
\newcommand{\nparams}{78{,}964}
\newcommand{\ntrain}{12{,}136}

\newcommand{\winSealedSP}{0.068}

\newcommand{\winSealedTthree}{0.049}

\newcommand{\nSealed}{1{,}210}
\newcommand{\winSealed}{0.061}
\newcommand{\leSealed}{74}
\newcommand{\ngapSealed}{605}
\newcommand{\gapSealed}{0.094}

\newcommand{\nBigSP}{575}
\newcommand{\winBigSP}{0.066}
\newcommand{\leBigSP}{69}
\newcommand{\ngapBigSP}{541}
\newcommand{\gapBigSP}{0.112}

\newcommand{\gapBigDTthree}{0.161}

\newcommand{\nBigD}{471}
\newcommand{\winBigD}{0.073}
\newcommand{\leBigD}{64}
\newcommand{\ngapBigD}{350}
\newcommand{\gapBigD}{0.136}

\newcommand{\winDten}{0.057}
\newcommand{\nchemDten}{555}
\newcommand{\leDten}{76}
\newcommand{\gapDten}{0.109}

\newcommand{\ngapDten}{226}

\newcommand{\gapstableDten}{0.105}

\newcommand{\winTtwo}{0.058}
\newcommand{\nchemTtwo}{589}
\newcommand{\leTtwo}{76}
\newcommand{\gapTtwo}{0.131}

\newcommand{\ngapTtwo}{196}

\newcommand{\gapstableTtwo}{0.122}

\newcommand{\winTthree}{0.050}
\newcommand{\nchemTthree}{423}
\newcommand{\leTthree}{81}
\newcommand{\gapTthree}{0.102}

\newcommand{\ngapTthree}{82}

\newcommand{\gapstableTthree}{0.100}

\newcommand{\winTfour}{0.056}
\newcommand{\nchemTfour}{177}
\newcommand{\leTfour}{82}
\newcommand{\gapTfour}{0.103}

\newcommand{\ngapTfour}{14}

\newcommand{\winSP}{0.065}
\newcommand{\nchemSP}{639}
\newcommand{\leSP}{71}
\newcommand{\gapSP}{0.084}

\newcommand{\ngapSP}{466}

\newcommand{\ratioLo}{0.76}
\newcommand{\ratioHi}{1.38}

\newcommand{\nchemAll}{2{,}050}
\newcommand{\leAll}{74}
\newcommand{\winAll}{0.061}
\newcommand{\gapAll}{0.102}
\newcommand{\ngapAll}{964}
\newcommand{\gapmeanAll}{0.16}
\newcommand{\nproto}{953}
\newcommand{\winProto}{0.076}
\newcommand{\gapProto}{0.110}
\newcommand{\ngapProto}{377}
\newcommand{\leProto}{62}

\title{\textbf{Learning Electronic Band Structures Across the \textit{s}, \textit{p}, and \textit{d} Blocks with an Empirical Pseudopotential Network}}
\author{Qiang Zhu}
\affil{Department of Mechanical Engineering and Engineering Science, University of North Carolina at Charlotte, Charlotte, NC 28223, USA. \texttt{qzhu8@charlotte.edu}}
\date{}

\hypersetup{pdftitle={Learning Electronic Band Structures Across the s, p, and d Blocks with an Empirical Pseudopotential Network},pdfauthor={Qiang Zhu}}
\let\SItopfraction\topfraction\let\SIbottomfraction\bottomfraction
\let\SItextfraction\textfraction\let\SIfloatpagefraction\floatpagefraction
\renewcommand{\topfraction}{0.95}\renewcommand{\bottomfraction}{0.95}
\renewcommand{\textfraction}{0.05}\renewcommand{\floatpagefraction}{0.75}
\begin{document}
\maketitle

\begin{abstract}
\noindent
Electronic band structures govern how crystals conduct electrons and absorb light. Density-functional theory (DFT) computes them through costly self-consistent iterations, while existing machine-learning models either predict isolated scalar properties or output Hamiltonians that still require DFT solvers. Here we present EpNet, a learned empirical pseudopotential for crystals spanning 63 elements across the \textit{s}, \textit{p} and \textit{d} blocks. An equivariant neural network builds its Hamiltonian from physics-motivated local potentials and nonlocal projectors. Trained solely on band energies, EpNet predicts complete band structures directly from atomic coordinates. On unseen chemical systems, its median errors are \winAll\ eV for band energies and \gapAll\ eV for band gaps. On 600,000 structures from an independent database, its band-gap errors are overall less than half those of existing models, at about 1\% of DFT cost. EpNet thus enables fast screening of electronic properties for materials discovery and electronic-structure simulations of large systems.
\end{abstract}

\section*{Introduction}

The electronic band structure of a crystal is critical for technological applications because it determines the band gap, carrier masses, Fermi surface, and optical response, serving as the starting point for transport, defect, and electron–phonon calculations in semiconducting and superconducting devices. Density-functional theory (DFT)\cite{Hohenberg1964,Kohn1965}  is the standard method for computing band structures and has been widely adopted to construct online databases providing public access to precomputed band gaps for millions of structures\cite{Jain2013,Curtarolo2012,Kirklin2015,Choudhary2020JARVIS,Haastrup2018C2DB,Draxl2019NOMAD,Schmidt2023Alexandria}. However, modern materials design demands rapid screening across vast chemical spaces consisting of several million to billion candidate structures, where standard DFT calculations become a severe bottleneck. Furthermore, DFT is computationally prohibitive for large-scale applications such as disordered systems, finite-temperature dynamics, or multi-body interactions like electron–phonon coupling.

Machine learning has emerged as a powerful tool to accelerate electronic structure calculations\cite{Kulik2022Roadmap}. Direct property-prediction models use equivariant graph neural networks to map crystal structures to target quantities like band gaps or densities of states\cite{Xie2018,Chen2019,Chen2021MultiFidelity,Choudhary2021ALIGNN,Ruff2024coGN}, reaching mean gap errors as low as 0.16 eV\cite{Ruff2024coGN,Dunn2020Matbench}. However, because they lack an explicit Hamiltonian formulation, predicted properties need not be mutually consistent, and derived features like carrier effective masses or gap nature cannot be extracted without training on separate labels.
To preserve the operator structure, machine-learned DFT Hamiltonians have been developed to predict converged Kohn–Sham matrices using localized atomic orbitals\cite{Schutt2019,Li2022DeepH,Gong2023DeepHE3,Zhong2023HamGNN,Gu2024DeePTB,Qian2026MACEH}, with universal implementations spanning the periodic table at meV-level matrix-element errors\cite{Wang2024DeepHUMM,Zhong2024UniHamGNN,Yin2026NextHAM}. Nevertheless, these models remain bound to specific DFT codes and basis sets. Extending them to higher-level electronic structure methods requires generating new matrix training sets\cite{Tang2024DeepHhybrid}, plane-wave representations disrupt network locality\cite{Gong2024DeepHPW}, and actual band-energy errors are often 10 to 50 times larger than the reported matrix-element errors\cite{Qian2026MACEH}. Models trained on computed properties rather than on matrices avoid the need for higher-level matrices. For molecules, an equivariant network that predicts an effective electronic Hamiltonian, trained on CCSD(T) data, proves to be more accurate than hybrid density functionals for a range of properties\cite{Tang2025MEHnet}.
On the other hand, semi-empirical tight-binding models eliminate the need for a DFT solver\cite{Cui2024PTBP,Garrity2023TB3,Froitzheim2025gxTB}, but their minimal basis set degrades conduction-band quality, leading to average band-gap errors of 0.7–0.8\,eV\cite{Park2025CHIPSTB}. Machine-learned tight-binding variants resolve this accuracy gap but remain limited to system-specific parameterizations\cite{Gu2024DeePTB}.

The empirical pseudopotential method (EPM)\cite{Cohen1966,Chelikowsky1976,CohenChelikowsky1988} combines several of these strengths. Assembled from atomic form factors in a plane-wave basis, an EPM Hamiltonian yields complete band structures at any $\mathbf{k}$-point through a single matrix diagonalization while bypassing self-consistent iterations in common DFT solvers. Although traditional EPM suffered from poor transferability due to material-by-material fitting, machine-learning models have recently revived the framework by predicting screened local potentials from DFT potentials\cite{KimSon2024} or band energies\cite{Kang2025MLEPM,Lin2025,ZhuEpNetSi}. However, these models remain limited to single elements or narrow semiconductor families.
Scaling EPM across the periodic table presents a fundamental physical challenge: while main-group $s$ and $p$ states are extended and well-described by weak, smooth pseudopotentials, transition-metal $d$ states form narrow resonances governed by localized orbitals\cite{Heine1967,Pettifor1970,HarrisonFroyen1980}. Historically, empirical approaches captured $d$ states only through strongly attractive $l=2$ terms fitted to individual metals\cite{FongCohen1970}, or by importing nonlocal operators and overlap matrices directly from ab initio pseudopotentials\cite{WangZunger1995,FuZunger1997,Cardenas2012AEP,Paudel2026TMD}. 

To address this challenge, we present EpNet, a unified learned EPM framework for crystals spanning 63 elements across the $s$, $p$, and $d$ blocks. EpNet assigns each atom a local potential derived from its chemical identity and local environment, alongside nonlocal channels matched to its valence states. Parameterized by \nparams\ weights and trained end-to-end through the eigensolver on band energies alone, EpNet predicts complete band structures directly from crystal coordinates. On held-out chemical systems, it achieves median errors of \winAll\,eV for band energies and \gapAll\,eV for band gaps. Evaluated on 600,000 completely unseen structures from the Alexandria database, EpNet achieves a median band-gap error of 0.11\,eV—overall less than half that of existing models—while reducing computational cost by two to three orders of magnitude relative to plane-wave DFT. To our knowledge, EpNet is the first learned EPM model capable of predicting complete band structures across more than 60 elements from crystal geometry alone. In the following sections, we detail the architecture and design of EpNet, evaluate its performance across benchmark datasets, discuss its downstream applications, and present our conclusions.

\section*{Methodology and design}

\begin{figure}[tbp]
\centering
\includegraphics[width=\textwidth]{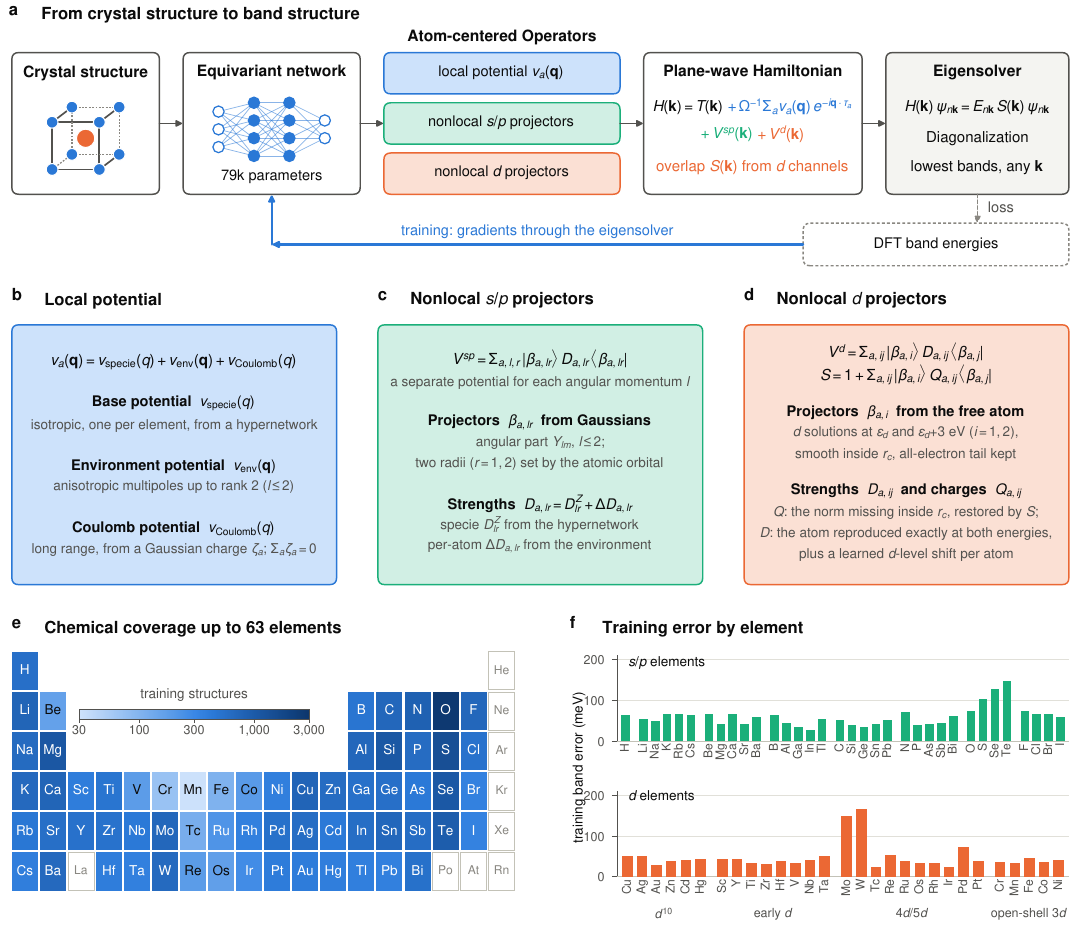}
\caption{\textbf{Design of EpNet.}
\textbf{a}, An equivariant network maps element features and atomic environments to atom-centered operators, which are assembled into a plane-wave Hamiltonian $H(\mathbf k)$ and overlap $S(\mathbf k)$, followed by a diagonalization. Training propagates the error in the PBE band energies of about 12,000 Materials Project crystals through the eigensolver.
\textbf{b}, Local potential: $v_{\rm specie}$ starts from a screened empty core (\textit{s}/\textit{p}) or the atom's own potential (\textit{d}).
\textbf{c}, Nonlocal \textit{s}/\textit{p} projectors.
\textbf{d}, Nonlocal \textit{d} projectors, built from the scalar-relativistic all-electron atom on its own local potential.
\textbf{e}, The 63 elements, shaded by the number of training structures that contain each; gray, outside the scope.
\textbf{f}, Training band error by element.}%
\label{fig:design}
\end{figure}

In the classical EPM, a crystal structure is mapped to a plane-wave Hamiltonian matrix (Fig.~\ref{fig:design}a).
For a unit cell of volume $\Omega$ with atomic positions $\bm\tau_a$, the Hamiltonian matrix elements in the plane-wave basis $|\mathbf k+\mathbf G\rangle$ are given by
\begin{equation}
H_{\mathbf G\mathbf G'}(\mathbf k)=\frac{\hb^2}{2m}|\mathbf k+\mathbf G|^2\delta_{\mathbf G\mathbf G'}
+\frac{1}{\Omega}\sum_a v_a(\Delta\mathbf G)\,e^{-i\Delta\mathbf G\cdot\bm\tau_a}+V^{\rm nl}_{\mathbf G\mathbf G'}(\mathbf k),
\qquad H\mathbf c=E\,S\,\mathbf c,
\label{eq:H}
\end{equation}
where $\Delta\mathbf G=\mathbf G-\mathbf G'$.
The overlap matrix $S$ is the identity matrix, except in cells that contain $d$-block atoms, as described below.

\paragraph{Local potential across the periodic table.}
To make the EPM transferable across chemical environments, we express the local potential $v_a(\mathbf q)$ of atom $a$ as summarized in Fig.~\ref{fig:design}b:
\begin{equation}
\begin{gathered}
v_a(\mathbf q)=v_{\rm specie}(|\mathbf q|)+v_{\rm env}(\mathbf q)+v_{\rm Coulomb}(|\mathbf q|),\\
v_{\rm env}(\mathbf q)=\sum_{\nu=1}^{4}\pi_{a,\nu}\,v^{\rm res}_{a,\nu}(\mathbf q),\qquad
v_{\rm Coulomb}(|\mathbf q|)=-\frac{4\pi e^2\zeta_a}{|\mathbf q|^2}\,e^{-\sigma^2|\mathbf q|^2/2}.
\end{gathered}
\label{eq:va}
\end{equation}

This formulation comprises three main terms.
First, the atomic specie potential $v_{\rm specie}(|\mathbf q|)$ defines an isotropic base potential parameterized by elemental features (e.g., period, group, electronegativity, and covalent radius). For $s$/$p$ elements, it starts from a Thomas--Fermi-screened empty-core potential\cite{Ashcroft1966}; for $d$ elements, it starts from the neutral atom's self-consistent potential. Second, the environmental correction $v_{\rm env}(\mathbf q)$ accounts for local coordination and chemical bonding. An equivariant neural network with a 6\,\AA\ cutoff predicts invariant, vector, and rank-2 tensor features for each atom, from which four multipole readouts $v^{\rm res}_{a,\nu}(\mathbf q)$ ($\nu=1,\dots,4$, each with $l\le2$ components) are constructed. A router mixes these readouts with per-atom weights $\pi_{a,\nu}$ (a softmax, $\sum_\nu\pi_{a,\nu}=1$) computed from the local valence-electron density and the electronegativity contrast relative to neighboring atoms. This term injects smooth, anisotropic environment effects into the potential, ensuring that the predicted band energies remain smoothly differentiable with respect to atomic displacements. Third, a Gaussian charge distribution on each atom (constrained to maintain cell neutrality) captures long-range electrostatic differences $v_{\rm Coulomb}(|\mathbf q|)$ between cations and anions, ensuring accurate band gaps for ionic compounds.

In addition, two design choices ensure that $v_a(\mathbf q)$ behaves as a true pseudopotential rather than a basis-set fit: (i) the plane-wave kinetic energy cutoff is element-dependent (12\,Ry by default, 16\,Ry for B and C, and 20\,Ry for N, O, F, and $d$-block elements) to accommodate compact orbitals, and (ii) band energies are explicitly constrained to remain convergent as the basis size is enlarged during the $s$/$p$ training stage. 

\paragraph{Nonlocal channels.}
The nonlocal potential $V^{\rm nl}_{\mathbf G\mathbf G'}(\mathbf k)$ distinguishes between extended and localized valence states.
For $s$/$p$ atoms, the nonlocal operator (Fig.~\ref{fig:design}c) consists of Kleinman--Bylander projectors\cite{Kleinman1982} built from two Gaussian functions per angular momentum channel at radii set by the element's valence orbitals, while the specie-level strengths come from the hypernetwork, with environment-dependent corrections per atom. Each $d$-block element instead carries a channel constructed directly from its free atom (Fig.~\ref{fig:design}d) to capture the localized $d$ states.
From the scalar-relativistic atom\cite{vanLenthe1993}, we take the $(n-1)d$ level $\varepsilon_d$ and orbital $u_d(r)$, and replace the orbital inside a core radius $r_c$ by a smooth, nodeless pseudo-orbital $\phi(r)$.
Crucially, $\phi(r)$ is not renormalized. Beyond $r_c$ it is identical to the all-electron orbital, so its tail, which overlaps with neighboring atoms and therefore sets the $d$-band width, is exact. Inside $r_c$ the smooth $\phi(r)$ holds less norm than the true orbital, and this missing norm is restored by an overlap operator $S$.
The local potential is the neutral atom's own beyond a radius $r_{\rm loc}$, stored as a fixed table $v(q)$.
On this potential, the $d$ channel is an ultrasoft pseudopotential\cite{Vanderbilt1990,Hamann2013} with two reference energies, $\varepsilon_1=\varepsilon_d$ and $\varepsilon_2=\varepsilon_d+3$\,eV,
$V^{d}=\sum_{ij}|\beta_i\rangle D_{ij}\langle\beta_j|$ with overlap $S=1+\sum_{ij}|\beta_i\rangle Q_{ij}\langle\beta_j|$.
The projectors $\beta_i$ are dual to the pseudo-orbitals $\phi_i$ at the two energies ($\langle\beta_i|\phi_j\rangle=\delta_{ij}$), and the strengths are
\begin{equation}
D_{ij}=\langle\phi_i|\,\varepsilon_j-T-V_{\rm loc}\,|\phi_j\rangle+\varepsilon_jQ_{ij},\qquad
Q_{ij}=\langle u_i|u_j\rangle-\langle\phi_i|\phi_j\rangle,\qquad i,j\in\{1,2\},
\label{eq:D}
\end{equation}
where $u_i(r)$ are the all-electron radial orbitals at the two reference energies ($u_1=u_d$), the brackets are radial integrals over the core region, so that $D_{ij}$ and $Q_{ij}$ are $2\times2$ matrices of numbers, and $Q_{ij}$ restores the norm and overlaps that the pseudo-orbitals miss inside $r_c$ (Supplementary equations~S4--S9).
By construction, the pseudo-atom reproduces the all-electron $d$ level and orbital exactly at both reference energies and interpolates the scattering between them. On the other hand, the $s$ and $p$ channels of $d$-block elements each use a single reference energy, their own atomic level ($\varepsilon_s$ or $\varepsilon_p$).
Because these projectors are recomputed from the stored atomic states at every evaluation, the reference channels remain exact as the learned network parameters evolve.

In a crystal environment, the $d$ channel changes through two learned corrections: (i) a shift $\delta$ of the $d$ level, one learned value per element plus one per atom from its environment, introduced as $D_{ij}\to D_{ij}+\delta\,\langle u_i|u_j\rangle$, which shifts the atomic $d$ level by $\delta$ to first order without changing the projectors; and (ii) a learned, bounded radial stretch $\lambda$ of the pseudo-orbitals, $\phi_i(r)\to\lambda^{-1/2}\phi_i(r/\lambda)$, one per element, for the seven elements where it remains numerically well conditioned (Supplementary Section~3). In addition, the augmentation term $\Delta V_aQ_{ij}$ is added to the projector strengths, $D_{ij}\to D_{ij}+\Delta V_aQ_{ij}$, in which $\Delta V_a$ is the crystal's local potential at the atom, minus the atom's own, averaged over the augmentation charge, so that the $d$ level follows the local electrostatic potential of the surrounding crystal\cite{Vanderbilt1990}.

\paragraph{Model training.}
We considered a total of \ntrain\ PBE\cite{Perdew1996} band structures calculated along high-symmetry paths from the Materials Project\cite{Jain2013} for model training. The dataset comprises non-magnetic crystal structures containing up to 24 atomic sites without $+U$ corrections. In total, the dataset covers 63 elements, ranging from 31 compounds containing Mn to 2,674 compounds containing O (Fig.~\ref{fig:design}e). Unlike previous approaches\cite{Dunn2020Matbench}, our held-out validation and test sets are defined by chemical system and structure type rather than through random splitting. Entire chemical systems (all structures sharing a given set of elements) and distinct structure types are assigned to the held-out sets via hashing prior to training. This strategy prevents polymorphs or close structural analogues of test materials from appearing in the training set, avoiding the artificially inflated predictive performance on novel chemical domains typical of random splits\cite{Meredig2018}. The held-out sets are excluded from all training stages, and no held-out chemical system appears in the training split.

EpNet was trained end-to-end using \nparams\ parameters according to the architecture detailed in Fig.~S1. The loss function combines the band-energy error within an energy window around the band gap or Fermi level, a dedicated gap penalty, and a weakly weighted regularization term across all occupied bands. For gapped compounds featuring open-shell $d$-block elements, the loss function also incorporates target labels for the unoccupied $d$ manifold.
Training follows a curriculum learning strategy\cite{Bengio2009}, from simpler to harder systems, with each stage initialized from the weights of the preceding one. First, $s$/$p$-block compounds are trained to establish the universal local potential. Next, small unit cells representing distinct $d$-block subgroups are introduced sequentially ($d^{10}$ elements, early $d$-block, $4d$/$5d$ elements, and open-shell $3d$ elements). Finally, the dataset is expanded to include cells from 8 sites to 24 sites via two steps. The resulting median training errors range between 0.03\,eV and 0.17\,eV across the elements, as detailed in Fig.~\ref{fig:design}f.  The highest training errors are predominantly concentrated in Mo, W, S, Se, and Te, originating primarily from layered mixed Mo/W dichalcogenide structures containing large vacuum gaps in the periodic supercell setup\cite{Bassman2018}. Specifically, 38 of the 64 evaluated training structures containing Mo or W exhibit these elevated median errors of 0.18\,eV. After excluding these layered systems, the median training errors are reduced to 0.04\,eV for Mo and W, and 0.05–0.06\,eV for S, Se, and Te. Hence, these errors are specific to layered stacks with large vacuum gaps rather than to Mo, W or the chalcogens in general (Supplementary Section~6).

\section*{Results}

\begin{figure}[tbp]
\centering
\includegraphics[width=\textwidth]{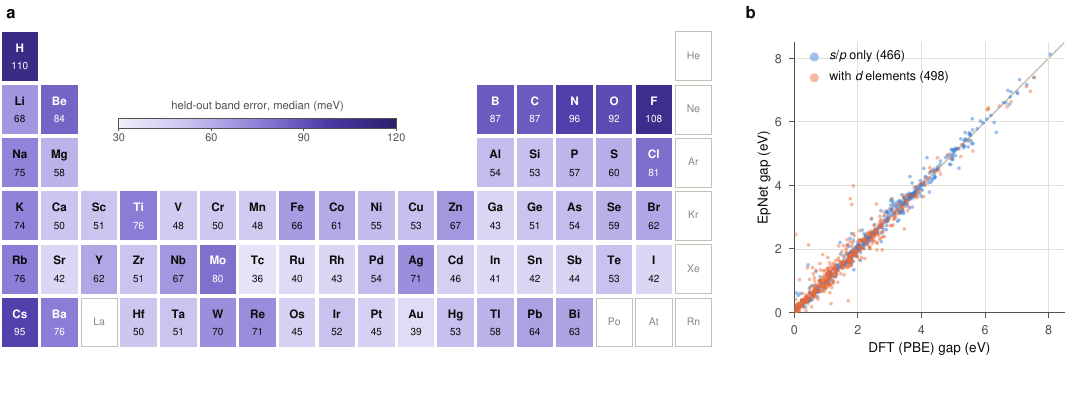}
\caption{\textbf{Accuracy on chemical systems absent from training.}
\textbf{a}, Median band error (meV) of the held-out structures that contain each element.
\textbf{b}, Band gaps of the \ngapAll\ gapped held-out structures.}%
\label{fig:accuracy}
\end{figure}

\paragraph{Model evaluation across 63 elements.}
Fig.~\ref{fig:accuracy} summarizes the performance of the final model on the \nchemAll\ held-out structures (containing up to 24 sites) whose chemical systems were completely absent from the training set. The median band-energy error within the evaluation window is \winAll\,eV, with \leAll\% of all structures falling within an error of 0.1\,eV. Representative band structures of held-out compounds, selected by error percentile ranging from the median to the 90th percentile, are presented in Fig.~S2.
When resolved by element (Fig.~\ref{fig:accuracy}a), the largest errors occur on the first-row elements H, F, N, O, and C (0.09–0.11\,eV). Across all held-out structures, the predicted band gaps yield a median error of \gapAll\,eV and a mean error of \gapmeanAll\,eV (Fig.~\ref{fig:accuracy}b), exhibiting consistent performance across diverse chemical domains. Evaluation on structure types absent from the training set and larger unit cells containing 25–60 sites shows slightly higher errors (Table~S1).
Overall, held-out and training errors remain closely aligned, with test-to-train error ratios ranging between \ratioLo\ and \ratioHi\ across the various $d$-block groups, confirming that the model avoids overfitting. Despite operating with a highly compact parameter space of \nparams\ weights, EpNet achieves a good balance between low model complexity and high predictive accuracy across the periodic table.

\begin{figure}[tbp]
\centering
\includegraphics[width=\textwidth]{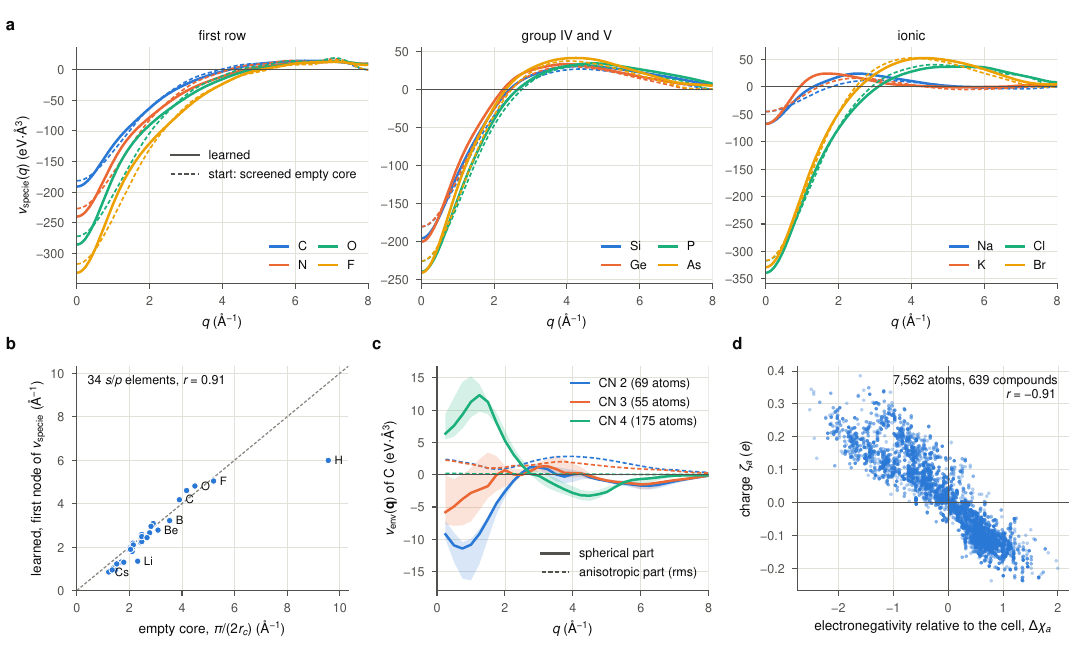}
\caption{\textbf{Learned local potentials of \textit{s}/\textit{p} elements.}
\textbf{a}, Base potentials $v_{\rm specie}(q)$ of first-row, group IV and V and ionic elements (solid) and the screened empty cores they started from (dashed).
\textbf{b}, First node of $v_{\rm specie}$ against that of the empty core, $\pi/(2r_c)$ with $r_c=0.53\,r_{\rm cov}$, for the 34 \textit{s}/\textit{p} elements.
\textbf{c}, Environment potential of the carbon atoms of scored \textit{s}/\textit{p} compounds, by coordination number: spherical part (median, solid; interquartile range, shading) and anisotropic part (root mean square over directions, dashed).
\textbf{d}, Charges $\zeta_a$ of the atoms of the held-out \textit{s}/\textit{p} compounds against their electronegativity relative to the cell mean.}
\label{fig:learned}
\end{figure}

\paragraph{Learned local potentials of \textit{s}/\textit{p} elements.}
We attribute much of this accuracy to the constraint that the learned potentials remain physical pseudopotentials in our design. Fig.~\ref{fig:learned}a shows that the base potentials of the \textit{s}/\textit{p} elements stay smooth, finite at $q=0$ as for screened ions and decaying by 8\,\AA$^{-1}$, and differ from the empty cores they started from mainly below 3\,\AA$^{-1}$.
The first node of a base potential is the wavevector at which $v_{\rm specie}(q)$ first changes sign, from attractive to repulsive, measuring the size of the ion core. For an empty core of radius $r_c$, it lies at $q=\pi/(2r_c)$.
Across all 34 \textit{s}/\textit{p} elements, the learned first node follows that of the starting empty core, with a Pearson correlation coefficient $r=0.91$ (Fig.~\ref{fig:learned}b).
Beyond these base potentials, the chemistry-specific part of the model lies in the environment potential and the charges.
The learned environment potential also follows bonding (see Fig.~\ref{fig:learned}c).
For carbon its spherical part shifts the potential near $q=1$\,\AA$^{-1}$ by about $\pm11$\,eV$\cdot$\AA$^3$, a tenth of the base potential there ($-123$\,eV$\cdot$\AA$^3$), in opposite directions for twofold and fourfold coordination.
Its anisotropic part is 1.6--1.8\,eV$\cdot$\AA$^3$ for twofold and threefold carbon but almost vanishes (0.13\,eV$\cdot$\AA$^3$) for fourfold carbon, as it should for tetrahedral sites, whose symmetry forbids dipole and quadrupole terms.
Fig.~\ref{fig:learned}d shows the learned atomic charges $\zeta_a$ of the held-out \textit{s}/\textit{p} compounds against their electronegativity relative to the cell mean. Their median magnitude is 0.10\,$e$, far below formal oxidation states, yet they correlate strongly with the atom's electronegativity ($r=-0.91$).
Switching them off in the trained model shows that on 240 ionic held-out \textit{s}/\textit{p} structures the median band error rises from 0.063 to 0.152\,eV, 77\% of the structures become worse by more than 0.01\,eV, and the median gap error of the 167 gapped ones rises from 0.077 to 0.393\,eV. In the meantime, the metals change less (0.044 to 0.091\,eV). This confirms that the charges are essential for ionic compounds, and that the model has learned to use them correctly.

\begin{figure}[tbp]
\centering
\includegraphics[width=\textwidth]{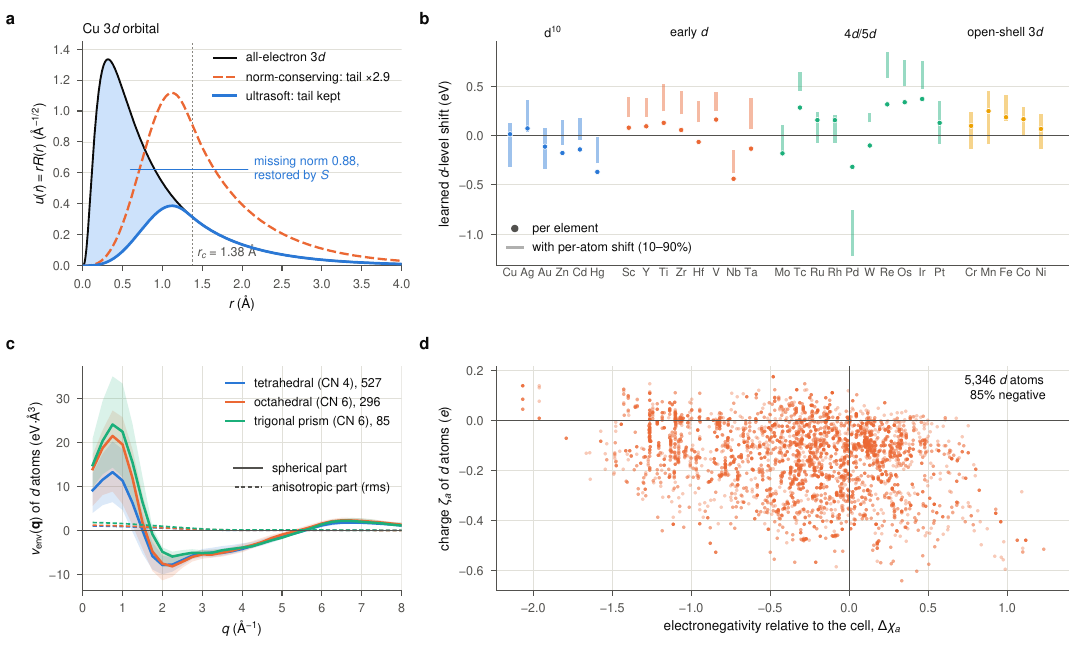}
\caption{\textbf{Learned potentials and \textit{d} channels of \textit{d} elements.}
\textbf{a}, Cu 3\textit{d} orbital: all-electron (black), the ultrasoft pseudo-orbital that keeps the all-electron tail (blue) and a norm-conserving pseudo-orbital obtained by renormalizing it (orange); shading: the norm restored by the overlap operator. 
\textbf{b}, Learned \textit{d}-level shifts: per element (dots) and, with the per-atom shift, the 10th--90th percentiles over the \textit{d} atoms of the scored compounds (bars).
\textbf{c}, Environment potential of \textit{d} atoms by coordination: spherical part (median, solid; interquartile range, shading) and anisotropic part (root mean square over directions, dashed).
\textbf{d}, Charges of the \textit{d} atoms of held-out compounds against their electronegativity relative to the cell mean.}
\label{fig:dchannel}
\end{figure}

\paragraph{Learned potentials of \textit{d} elements.}
For the \textit{d} elements, the essential physics is built in before any learning.
The width of a \textit{d} band is set by the overlap of \textit{d} orbitals on neighboring atoms, which lies in the orbital tail (Fig.~\ref{fig:dchannel}a).
EpNet therefore builds each \textit{d} channel from the free atom's \textit{d} orbital and keeps its all-electron tail unchanged; the norm missing inside the core radius is restored by the overlap operator, and the channel has no spurious bound states\cite{Gonze1991}.
Before any training, this construction places the \textit{d}-band widths of seven \textit{d}$^{10}$ compounds within 1.13--1.26 times PBE (Supplementary S3, Fig.~S3 and Table~S2).

After model training, the corrections learned on top of this construction are small, and most of them follow physical trends.
The base potentials stay within 0.2\% (root mean square) of their atomic tables, and the orbital stretch within 1\% of the atomic orbital.
The learned \textit{d}-level shifts (Fig.~\ref{fig:dchannel}b) are a few tenths of an eV, the size of the correction between a free neutral atom and a \textit{d} ion screened in a crystal.
They sort by shell filling.
Per element, they are slightly negative for the filled \textit{d}$^{10}$ shells (median $-0.13$\,eV) and positive for the open-shell 3\textit{d} and the 4\textit{d}/5\textit{d} elements (medians $+0.17$ and $+0.16$\,eV).
Per atom, they are largest on the most cationic sites.
The environment potential (Fig.~\ref{fig:dchannel}c) is more repulsive at more crowded sites, as the overlap of neighboring charge densities makes it: its spherical part near $q=1$\,\AA$^{-1}$ rises from 11\,eV$\cdot$\AA$^3$ at fourfold sites to 20--23, 29 and 37\,eV$\cdot$\AA$^3$ at six-, eight- and twelvefold sites.
Its anisotropic part obeys the crystal-field selection rules: it is small at tetrahedral and octahedral sites (0.9--1.1\,eV$\cdot$\AA$^3$), whose cubic symmetry forbids dipole and quadrupole terms, and larger at trigonal prisms (1.6\,eV$\cdot$\AA$^3$), where a quadrupole term splits $d_{z^2}$ from the in-plane $d_{xy}$ and $d_{x^2-y^2}$ orbitals.
However, the charges of the \textit{d} atoms are not physical (Fig.~\ref{fig:dchannel}d): 85\% of those in held-out compounds are negative (median $-0.12\,e$), and they correlate only weakly with electronegativity ($r=-0.25$, against $-0.91$ for \textit{s}/\textit{p} atoms).
A negative charge raises the potential of the atom, and the model appears to use it, with the mostly positive \textit{d}-level shifts, to lift a \textit{d} level that sits too deep; constraining the charges to be non-negative lost the fit of the \textit{d} gaps, so we read them as effective parameters.

\paragraph{Application to large-scale materials screening.}

We next apply EpNet to the large-scale Alexandria PBE database\cite{Schmidt2023Alexandria}, from which we extracted 600,617 non-magnetic structures that are completely absent from the Materials Project, span the 63 \textit{s}/\textit{p}/\textit{d} elements, contain up to 24 sites and lie within 0.1\,eV/atom of the convex hull.
Since Alexandria gives band gaps but not band structures, EpNet scored all of them based on the gap values and the metal/insulator classification. For a fair comparison with other methods, we also drew a sample of 5,000, 1,000 per group to score tight-binding models below.

\begin{figure}[tbp]
\centering
\includegraphics[width=\textwidth]{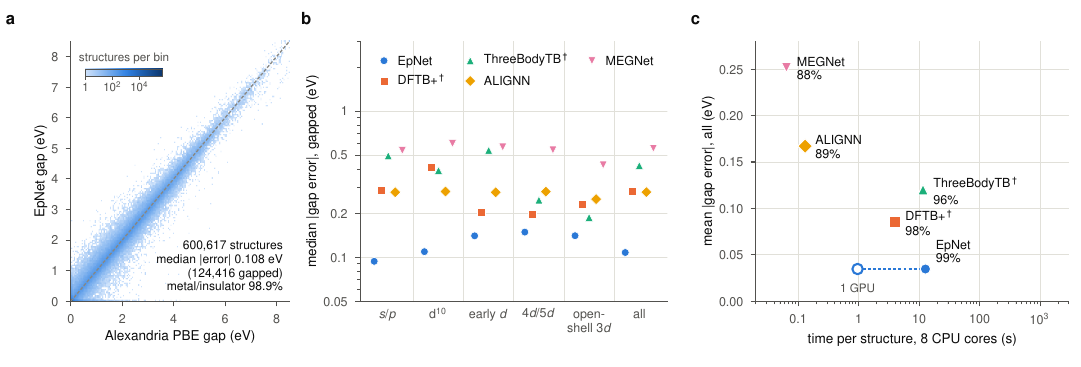}
\caption{\textbf{Benchmarking unseen Alexandria materials and comparison with other models.} Alexandria structures absent from the Materials Project: all 600,617 in \textbf{a}; in \textbf{b} and \textbf{c}, all of them for EpNet, ALIGNN and MEGNet and the 5,000 test structures for the two tight-binding models marked with $^\dagger$ (ThreeBodyTB on the 1,597 within its fit).
\textbf{a}, EpNet (screening mode) against Alexandria band gaps, structures per bin.
\textbf{b}, Median gap error of the gapped structures, by group.
\textbf{c}, Mean gap error over all structures against time per structure on 8 CPU cores; labels, metal/insulator agreement; open marker, EpNet on one GPU.}
\label{fig:alexandria}
\end{figure}

Fig.~\ref{fig:alexandria}a shows that over all 600,617 structures EpNet's median gap error is 0.108\,eV over the 124,416 gapped ones, 98.9\% are correctly classified as metal or insulator, and the distribution of gaps matches Alexandria's. %
On structures present in both databases, the Alexandria and Materials Project gaps differ by a median of only 0.03\,eV, so Alexandria is a fair external test for a model trained on Materials Project labels.
Fig.~\ref{fig:alexandria}b,c compares EpNet with existing models on the same structures.
Two graph networks trained on Materials Project PBE gaps, ALIGNN\cite{Choudhary2021ALIGNN} and MEGNet\cite{Chen2021MultiFidelity}, are fast to run on the whole pool, but they only yield the scalar gap values and their median gap errors are 0.279 and 0.555\,eV, 1.8--3.0 and 3.1--5.7 times EpNet's in the five groups, and they classify many metals as insulators (89\% and 88\%, against 99\% for EpNet). 
Two tight-binding models also return full band structures. On the 5,000-structure sample their median gap errors are 0.282\,eV (DFTB+\cite{Hourahine2020DFTBplus} with the PTBP parameters\cite{Cui2024PTBP}), against 0.105\,eV for EpNet on the same structures, and 0.424\,eV (ThreeBodyTB\cite{Garrity2023TB3}, on the 1,597 structures within its fit), against 0.091\,eV.
Hence, EpNet is the most accurate model in every group. %

In addition to its predictive accuracy, EpNet is compelling due to its low computational cost (Fig.~\ref{fig:alexandria}c). For gapped systems, EpNet requires approximately 12\,s per structure on 8 CPU cores, compared to 4\,s and 12\,s for tight-binding models on identical CPU hardware, and roughly 0.1\,s for graph neural networks, which yield only the band gap. However, GPU acceleration significantly reduces this runtime. Using EpNet, screening 614,729 structures of the Alexandria database required 18.5\,h across 16 GPUs, a median of 0.9\,s per structure on a single GPU.
We further benchmarked EpNet's computational cost against standard DFT solvers using two distinct test sets (Figs.~S4 and S5). On 24 held-out structures containing 2–24 atoms, computing a converged PBE band structure using Quantum ESPRESSO\cite{Giannozzi2009,Giannozzi2017} with SSSP pseudopotentials\cite{Prandini2018SSSP} required 3–984\,s. This corresponds to a median wall-clock time 99 times longer than EpNet on a single GPU, and 12 times longer than EpNet on the same 16 CPU cores (Fig.~S4). The speedup grows with cell size up to 24 atoms (Fig.~S4); for larger cells it becomes a constant factor, as both codes scale as $N^{2.2}$--$N^{2.6}$ (Fig.~S5).
On FCC supercells of Al and Cu containing more than 64 atoms, EpNet on a single GPU is 110–200 times faster for Al and 420–470 times faster for Cu per $\mathbf{k}$-point than Quantum ESPRESSO on 16 CPU cores, requiring only 10\,s and 82\,s per $\mathbf{k}$-point for 512-atom supercells of Al and Cu, respectively. EpNet's efficiency stems from bypassing the self-consistent field cycle, employing a significantly smaller plane-wave basis set (a 12–20\,Ry kinetic energy cutoff compared to 35–90\,Ry in standard DFT), and leveraging full GPU parallelization. %
Table~S3 also provides a detailed comparison of EpNet against several other machine-learning and semi-empirical electronic structure methods.

\section*{Conclusions}

We have demonstrated that a physics-based empirical pseudopotential network with \nparams\ parameters captures the electronic band structures of crystals across the $s$, $p$, and $d$ blocks of the periodic table. The central innovation is a Hamiltonian decomposed into distinct local and nonlocal operators. In this design, a universal local potential for all elements is learned from atomic features and local structural environments, while the nonlocal channels use learned projectors for the extended $s$ and $p$ states and projectors built from the free atom for the localized $d$ states. On chemical systems absent from the training set, EpNet, without self-consistent iterations or reliance on a DFT solver, predicts complete band structures directly from crystal coordinates with median errors of \winAll\,eV for band energies and \gapAll\,eV for band gaps. Crucially, benchmarks on an independent database show band-gap errors 2--5 times smaller than those of direct property-prediction networks and tight-binding models.

The remaining errors are concentrated within a few specific chemical regimes (Supplementary Section~6 and Fig.~S6). 
Furthermore, the model currently inherits the domain limits of its training labels. The present model is limited to PBE band energies of non-magnetic crystals without spin–orbit coupling. However, because EpNet relies solely on band energies for training, it can be readily extended or fine-tuned on hybrid-functional\cite{Tang2024DeepHhybrid,Adhikari2023HSE}, $GW$\cite{Knosgaard2022GW,Lin2025}, or experimental\cite{CohenChelikowsky1988} band data without requiring full Hamiltonian matrices. Looking forward, $f$-block elements can be incorporated using the same atomic construction established for $d$ channels, and the underlying Hamiltonian formulation can be extended to enable application to absolute band alignment and electron–phonon coupling calculations\cite{Giustino2017,Li2024DFPT,Zhong2024EPC,Lin2025,ZhuEpNetEPC}.

\section*{Methods}

\subsection*{Hamiltonian and basis}
The basis is centered on each $\mathbf k$, $|\mathbf k+\mathbf G|\le g_c$, and $\mathbf k$ is folded into the first Brillouin zone before $H(\mathbf k)$ is built\cite{ZhuEpNetSi}.
The kinetic-energy cutoff depends on the element, 12\,Ry by default.  
Compact 2\textit{p} states need more plane waves than heavier \textit{p} states, so we used 16\,Ry for B and C and 20\,Ry for N, O, F and all \textit{d} elements. Like other plane-wave codes, each cell uses the largest cutoff of its elements. %
In the first stage of training, on the \textit{s}/\textit{p} compounds, a basis-consistency term, the mean absolute difference of the bands at four $\mathbf k$ points between the cell's cutoff and 20\,Ry after removing a rigid shift, was added to the loss with weight 1 for one cell per GPU and step, so that the learned potential does not exploit the truncation of the basis.
The $\mathbf G=0$ component of the local potential is set to zero, which fixes the energy reference. Band energies are therefore compared after alignment.
For cells with \textit{d} atoms the generalized problem $H\mathbf c=ES\mathbf c$ is solved through the Cholesky factor of $S$ on the dense path, and through the exact low-rank $S^{-1/2}$ (L\"owdin form\cite{Lowdin1950}) on the iterative path.

\subsection*{Specie potentials}
\textbf{Base potentials.}
Each element is described by 15 features: a constant, $Z/100$, period/7, group/18, the \textit{s}, \textit{p} and \textit{d} valence occupancies, one-hot \textit{s}, \textit{p} and \textit{d} block labels, the Pauling electronegativity $\chi/4$, the covalent radius\cite{Cordero2008}/2.5\,\AA, $Z^2/10^4$, the model valence $Z_m/12$ and the number of model \textit{d} electrons, which separates valence \textit{d} shells (Zn--Hg) from the filled semicore shells of Ga--Bi.
A hypernetwork, a multilayer perceptron with one hidden layer of 32 units with sigmoid linear unit (SiLU) activations, maps these features to the specie projector strengths and to the coefficients of the specie potential,
\begin{equation}
v_{\rm specie}(q)=\sum_{n=0}^{11}c^{Z}_n\,\phi_n(q;8)+\sum_{n=0}^{25}t^{Z}_n\,\phi_n(q;17),\qquad \phi_n(q;q_c)=\tfrac12(1+\cos\pi x)\cos n\pi x,\quad x=\min(q/q_c,1),
\label{eq:vz}
\end{equation}
with $q_c$ in \AA$^{-1}$.
The second sum, a high-$q$ tail, is needed because the 20\,Ry basis couples plane waves up to $|\Delta\mathbf G|\approx17$\,\AA$^{-1}$.
The coefficients are the sum of the hypernetwork output, a small per-element residual with an $L_2$ penalty ($3\times10^{-6}$) and a fixed prior.
For \textit{s}/\textit{p} elements, the prior is a Thomas--Fermi-screened Ashcroft empty-core pseudopotential\cite{Ashcroft1966}.
It is set by the element's valence $Z_m$ and its core radius $r_c=0.53\,r_{\rm cov}$, a fixed multiple of the covalent radius $r_{\rm cov}$, since pseudopotential core radii scale linearly with atomic radii\cite{Zunger1981}.
Because the prior depends only on element features, it is defined for every element.
For \textit{d} elements it is the tabulated all-electron local potential described below, and the learned deviation from it enters scaled by 0.01.
The hypernetwork's output layer and the residuals start at zero.

\noindent\textbf{Environment terms and charges.}
We extended the environment encoder from the previous EpNet-Si\cite{ZhuEpNetSi} to many species, using neighbor moments up to $l=2$ within 6\,\AA\ under periodic boundary conditions, 8 radial functions, a single gated pass, and 64 channels of invariant ($h_a$), vector ($\mathbf d_a$) and traceless rank-2 ($\mathbf Q_a$) features.
The species embedding is a learned linear map of the element features plus a penalized per-species residual ($L_2$ weight $10^{-3}$), so an element enters through its features rather than through an independent vector.
The norms of $\mathbf d_a$ and $\mathbf Q_a$ that feed the gates are smoothed as $s(\sqrt{1+|\mathbf x|^2/s^2}-1)$ with $s=0.05$, which removes the cusp of $|\mathbf x|$ at high-symmetry sites, where $\mathbf d_a=\mathbf Q_a=0$, and keeps second derivatives of the bands with respect to atomic displacements well defined.
Four readouts map the encoder features to form factors with $l=0,1,2$ components and the $(-i)^l$ phases that keep $V(-\mathbf G)=V(\mathbf G)^*$, with an additional 12-term cosine basis in the monopole channel.
A router mixes them with per-atom softmax weights $\pi_{a,\nu}$ from a multilayer perceptron (one hidden layer of 16 units) on eight standardized inputs: the logarithm of the local valence-electron density $n_v(r)=\sum_jZ_{m,j}w(r_{aj};r)/(\tfrac43\pi r^3)$ at $r=2$, 3, 4, 5, 6, 8 and 10\,\AA, normalized by 0.2\,$e$/\AA$^3$, where $w$ is a smooth neighbor count, and the mean electronegativity difference to the neighbors within 4\,\AA.
A linear head on the invariant features gives raw charges whose cell mean is subtracted, so that $\sum_a\zeta_a=0$ holds exactly, while the charges enter only through the Coulomb potential $v_{\rm Coulomb}$ of equation~\eqref{eq:va} ($\sigma=0.5$\,\AA), and the head starts at zero.

\subsection*{Nonlocal projectors} 
\textbf{\textit{s}/\textit{p} elements}.
The nonlocal term of equation~\eqref{eq:H} is $V^{\rm nl}=V^{sp}+V^{d}$, and the first term is
\begin{equation}
V^{sp}_{\mathbf G\mathbf G'}(\mathbf k)=\sum_{a\in sp}\sum_{l\le l_{\max}(Z_a)}\sum_{r=1}^{2}\big(D^{Z_a}_{lr}+\Delta D_{a,lr}\big)\sum_{m}\beta^{lm}_{ar}(\mathbf p)\,\beta^{lm*}_{ar}(\mathbf p'),
\label{eq:nl}
\end{equation}
with $\mathbf p=\mathbf k+\mathbf G$, $\mathbf p'=\mathbf k+\mathbf G'$ and Gaussian radial projectors carrying the angular factor of channel $l$\cite{Kleinman1982}.
The two radial functions of channel $l$ have real-space radii of 1.2 and 0.7 times $r_l$, the peak radius of the element's valence \textit{s} or \textit{p} orbital in the neutral all-electron atom computed in the local-density approximation (LDA)\cite{Perdew1981PZ}, multiplied by a learned global factor per radial function and a learned factor per element and channel.
First-row atoms (B--F) have \textit{s} and \textit{p} projectors only, because a \textit{d}-like channel at their compact 2\textit{p} radius produced spurious states in development; for Ca, Sr, Ba and Cs the $l=2$ channel sits at the radius of the lowest empty $(n-1)d$ orbital, which forms their conduction-band edge, and for the other \textit{s}/\textit{p} elements at the \textit{p} radius.
The specie strengths $D^Z_{lr}$ come from the hypernetwork and the per-atom corrections $\Delta D_{a,lr}$ from a zero-initialized multilayer perceptron (one hidden layer of 16 units) on the 192 encoder invariants and two coordination numbers; one strength per $(l,r)$ multiplet keeps the projector rotationally invariant.

\noindent\textbf{\textit{d} elements}. 
Unlike \textit{s/p} elements, each \textit{d} element is solved as a neutral, spherical, scalar-relativistic (zeroth-order regular approximation\cite{vanLenthe1993}) LDA atom in its experimental ground-state configuration; valence eigenvalues agree with the NIST scalar-relativistic LDA tables\cite{Kotochigova1997,NIST141} to within 0.041\,eV.
Inside $r_c$ the orbital is $r^{l+1}(a+br^2+cr^4)$, matched to the all-electron orbital in value, slope and curvature; $r_c$ is the smallest radius for which the plane-wave components above 20\,Ry carry less than 0.05\,eV of kinetic energy (1.14--1.38\,\AA\ for \textit{d}, 0.77\,\AA\ for Y; Table~S2).
The orbital is not renormalized; the norm deficits $q_d$ range from 0.02 (Y) to 0.93 (Zn).
Beyond $r_{\rm loc}$ (1.2--1.8\,\AA) the local potential equals the neutral atom's self-consistent potential and inside it is a smooth polynomial matched in value, slope and curvature.
The neutral-atom potential extends to 5--6\,\AA, farther than the cosine basis of the learned potentials can represent.
It therefore enters as a fixed table of $v(q)$ (1,601 points up to 24.7\,\AA$^{-1}$), which reproduces the potential in real space to 0.003\,eV.
The radius $r_{\rm loc}$ is the smallest at which every channel passes the checks described below.
The \textit{s} and \textit{p} channels use one reference energy and the \textit{d} channel two, $\varepsilon_d$ and $\varepsilon_d+3$\,eV ($+6$\,eV for Sc), the second from the all-electron solution at that energy pseudized at the same $r_c$ (Supplementary equations~S7--S9).
Several choices keep the construction numerically stable.
The second orbital is normalized within the core, and $Q$ is replaced by its positive-semidefinite part, a change below 0.003.
The learned \textit{d}-level shifts enter through $D$ rather than through $\chi$, so that they cannot drive $B$ toward singularity.
The orbital stretch $\phi(r)\to\lambda^{-1/2}\phi(r/\lambda)$ is learned only for the seven elements whose operator norms stay within three times their atomic values over the allowed range (Sc, Y, Nb, Mo, Hf, Ta, W).
The \textit{p} channel lies close to its Kleinman--Bylander singular point and is not learned.
The pseudo-atom of every \textit{d} element, solved at 20\,Ry, is tested in five ways: (i) it reproduces the all-electron levels within 0.04\,eV; (ii) its overlap is positive definite ($S_{\min}\ge0.92$); (iii) no channel has a bound state below its reference level; (iv) its \textit{d} tail is within 2\% of the all-electron orbital (at 30\,Ry); and (v) its \textit{s} and \textit{p} scattering errors are 0.01--0.20\,eV from 4\,eV below to 6\,eV above the reference levels, except for the diffuse 5\textit{p} channels of Y and Zr (0.23--0.53\,eV).
We accept these two exceptions because the simpler local potential they would otherwise use gives far larger errors in crystals. %
The channels enter the projector slots of \textit{d} atoms; the Gaussian $l=2$ projector used for \textit{s}/\textit{p} atoms is switched off for them, since in single-crystal fits it allowed the fit to expel the \textit{d} state.
The augmentation term adds $\Delta V_aQ_{ij}$ to $D_{ij}$, with $\Delta V_a$ the cell's local potential, minus the atom's own, averaged over the augmentation density $u^2-\phi^2$.
A per-atom \textit{d}-level shift $\delta_a$ comes from a small network (one hidden layer of 16 units, zero-initialized output) on the atom's invariant features.

\subsection*{Valence electron condition, data and splits}
The model describes valence electrons only: $Z_m$ counts the valence \textit{s} and \textit{p} electrons of \textit{s}/\textit{p} elements and the $ns+(n-1)d$ electrons of \textit{d} elements (for example Ti: 4, Fe: 8, Pt: 10).
For each reference structure, the bands of the semicore shells that the Materials Project datasets treat as valence are counted from their valences and removed from the bottom of the spectrum.
Structures with a non-integer count, or with a removed band within 4\,eV of the band edge or Fermi level, are rejected.
In the retained structures, every removed band lies at least 4.5\,eV below the band edge or Fermi level, so the bands near the gap are untouched.
Every gapped \textit{d} structure confirmed the valence table.
The data are the PBE line-mode band structures of the Materials Project computed with VASP\cite{Kresse1996} along standard paths\cite{Setyawan2010}.
We exclude spin-polarized or $+U$ calculations and the \textit{f} block systems, which leaves 22,553 structures of up to 24 sites (consisting of 8,464 with only \textit{s}/\textit{p} elements and 14,089 with \textit{d} elements).
Structures are assigned to splits by hashing first their chemical system and then their prototype.
The \textit{s}/\textit{p} structures were assigned first, and their splits were kept when the \textit{d} structures were added.
Within the training split, 20\% of the chemical systems are held out from the loss as an inner validation set and 10\% as a checkpoint-selection set.

\subsection*{Loss, eigensolver and training}
The loss of a structure is the mean absolute error of its band energies within $\pm2$\,eV of midgap (gapped structures, aligned at the valence-band maximum) or of the Fermi level (metals), with the band-edge bands weighted by 5, plus the absolute error of the predicted gap and an all-occupied-band term with weight 0.3.
For gapped structures with open-shell \textit{d} atoms the labels include the lowest $5N_d$ conduction bands, the empty \textit{d} manifold, and the window extends to their top.
Only the lowest $n_{\rm occ}+n_{\rm emp}$ bands are computed: each structure keeps its eigenvectors from the previous visit, which seed Chebyshev-filtered subspace iteration\cite{Zhou2006}, and above 3,000 plane waves the Hamiltonian is never formed, the local potential being applied on a fast Fourier transform (FFT) grid and the nonlocal and overlap terms in low-rank form.
Gradients with respect to a model parameter $\theta$ follow from the Hellmann--Feynman theorem\cite{Feynman1939} in its generalized form,
\begin{equation}
\frac{\partial E_{n\mathbf k}}{\partial\theta}=\mathbf c_{n\mathbf k}^\dagger\left(\frac{\partial H(\mathbf k)}{\partial\theta}-E_{n\mathbf k}\,\frac{\partial S(\mathbf k)}{\partial\theta}\right)\mathbf c_{n\mathbf k},\qquad \mathbf c_{n\mathbf k}^\dagger S(\mathbf k)\,\mathbf c_{n\mathbf k}=1.
\label{eq:hf}
\end{equation}
Training uses AdamW\cite{Loshchilov2019} without weight decay, batches of 16 structures and a cosine learning-rate decay, with separate rates for the species parameters.
Training began with the \textit{s}/\textit{p} compounds: 17,800 steps from scratch on structures of up to 24 sites.
Three warm-start stages followed (1,500, 2,500 and 2,500 steps), which added a high-$q$ tail to the specie potentials and 596 structures of 25--60 sites rich in N and C--H.
Throughout these \textit{s}/\textit{p} stages, a penalty kept each cell's bands nearly unchanged when its basis was raised to 20\,Ry.
The \textit{d} elements were then added to the species table and trained in stages, each starting from the previous one and each with criteria fixed before it ran (Supplementary Information).
First, cells of up to 3 sites introduced the \textit{d} groups one at a time: \textit{d}$^{10}$, early \textit{d}, 4\textit{d}/5\textit{d} and open-shell 3\textit{d} elements, each stage with \textit{s}/\textit{p} cells of up to 8 sites as anchors (3,000 steps per stage, learning rate $10^{-3}$).
Next, all cells of up to 8 sites were trained for 22,000 steps (learning rates $10^{-3}$, then $10^{-4}$).
Then all \ntrain\ training cells of up to 24 sites and 8,700 plane waves were trained for 11,500 steps (learning rate $10^{-4}$), with the new cells above 8 sites weighted by 4.
A final 22,000 steps on the same cells used half the learning rate, with the \textit{s}/\textit{p} cells weighted by 1.5 and the new cells by 1.
The basis-consistency penalty of the first stage was not used in these stages.
Each structure contributes a subset of its path that always includes the path vertices and the band extrema (median 48 $\mathbf k$ points for cells of up to 8 sites and 33 for larger cells; 32 and 16 in the final stage).
The final stage took 17\,h on four NVIDIA H100 GPUs for one model and 35\,h on four L40S GPUs for the other.

\subsection*{Evaluation}
Structures are scored on low-resolution band paths with 64 $\mathbf k$ points. %
The band error is the mean absolute error within $\pm2$\,eV of midgap (gapped) or the Fermi level (metals, after the median shift within that window); the gap error is the difference of the path gaps.
Alexandria PBE (release 2025.07.02, 5.78 million entries) was filtered to the training scope, which leaves 615,564 structures, of which 600,617 match no in-scope Materials Project structure according to pymatgen's structure matcher\cite{Ong2013}.
From these, 1,000 per group were drawn, and the sample and metrics were fixed before scoring.
EpNet's gap is computed over a standard band path\cite{Setyawan2010} and a $\Gamma$-centered mesh with 0.12\,\AA$^{-1}$ spacing, while a structure is metallic when the gap closes or the electron count is odd.

\noindent
\textbf{Setting of other models}: For the benchmark of Alexandria data, we also employed (i) MEGNet; (ii) ALIGNN models trained on Materials Project PBE gaps; (iii) DFTB+ 25.1 with the complete PTBP parameter set; and (iv) ThreeBodyTB 1.1.1 (fitted to PBEsol). %
These models ran on 8 CPU cores (ALIGNN and MEGNet in batches, their total time divided by the number of structures), while EpNet was run on both 8 CPU cores and one GPU. The comparison with the plane-wave DFT code is described in Supplementary Section~4 (Figs.~S4 and S5).

\noindent
\textbf{EpNet band structure screen mode.} 
For screening, EpNet takes the gap from the band path alone, thinned to every fourth point (every third for the first 28,000 structures of the screen) with all vertices kept, and computes the bands at successive $\mathbf k$ points by Chebyshev-filtered subspace iteration started from the previous point's eigenvectors with a tolerance of 5\,meV.
The screen covered all in-scope structures except the 835 that match sealed Materials Project structures, 614,729 in total, in 285 GPU-hours (18.5\,h on up to 16 GPUs: A40, A100 and H100). %

\subsection*{Data and code availability}
The code is available at \url{https://github.com/MaterSim/EpNet-spd} under the MIT license, together with the trained models (both seeds and every checkpoint of the training curriculum), the recorded arguments of every training stage, the training manifest with its splits, the atomic \textit{d} channels and the per-structure scores of the final model.
The training labels are Materials Project\cite{Jain2013} band structures, which the repository's scripts download, and the external test uses the Alexandria database\cite{Schmidt2023Alexandria}.
The web application at \url{https://mmi.charlotte.edu/epnet} gives access to the final model.

\section*{Acknowledgments}
This research was sponsored by the U.S. Department of Energy, Office of Science, Office of Basic Energy Sciences, and the Established Program to Stimulate Competitive Research (EPSCoR) under the DOE Early Career Award No. DE-SC0024866. Computations were performed on the resources of University Research Computing at the University of North Carolina at Charlotte.

\section*{Competing interests}
The author declares no competing interests.

\bibliographystyle{unsrtnat}
\bibliography{refs}

\clearpage
\let\topfraction\SItopfraction\let\bottomfraction\SIbottomfraction
\let\textfraction\SItextfraction\let\floatpagefraction\SIfloatpagefraction
\setcounter{figure}{0}\setcounter{table}{0}\setcounter{equation}{0}
\renewcommand{\figurename}{Fig.}
\renewcommand{\thefigure}{S\arabic{figure}}\renewcommand{\theHfigure}{S\arabic{figure}}
\renewcommand{\thetable}{S\arabic{table}}\renewcommand{\theHtable}{S\arabic{table}}
\renewcommand{\theequation}{S\arabic{equation}}\renewcommand{\theHequation}{S\arabic{equation}}
\begin{center}
{\Large\bfseries Supplementary Information}
\end{center}

\FloatBarrier
\section*{1. The network, the Hamiltonian and the training loss}

Fig.~S1 shows the complete model from the crystal structure to the band energies, including the element features, the router and the equivariant encoder, the operators they produce, and the assembly of the plane-wave Hamiltonian. 

\begin{figure}[htbp]
\centering
\includegraphics[width=\textwidth]{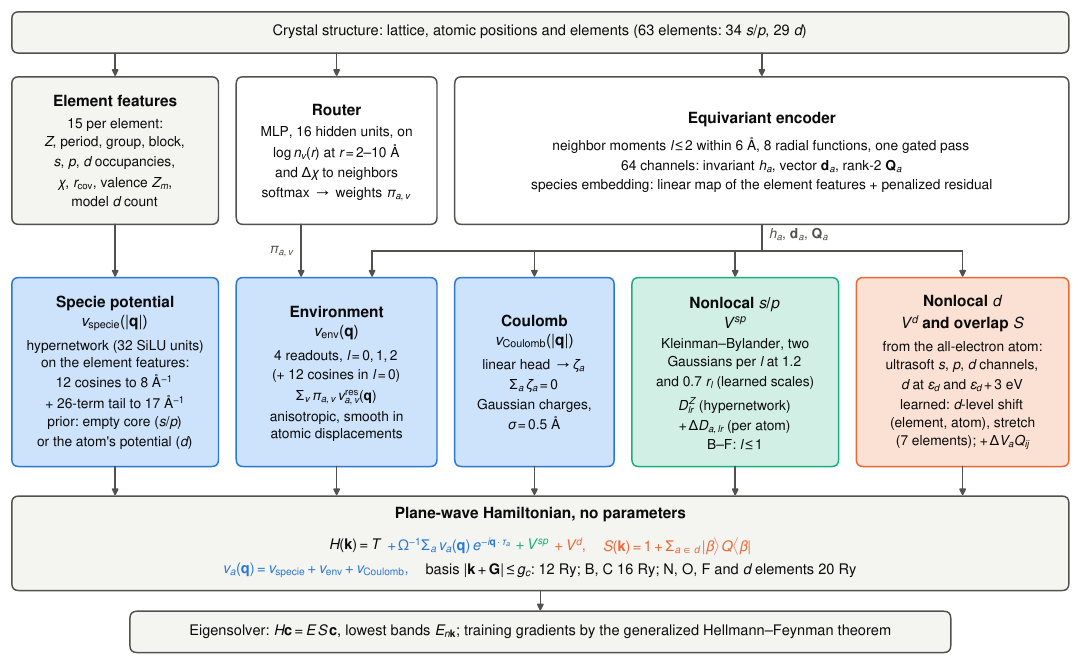}
\caption{\textbf{The full network and the assembly of the Hamiltonian.} Gray: fixed inputs and operations; white: learned networks; blue, aqua and orange: the local potential and charges, the \textit{s}/\textit{p} projectors and the \textit{d} channels, as in main-text Fig.~1. The encoder features $h_a$, $\mathbf d_a$ and $\mathbf Q_a$ feed the environment term, the charges and the per-atom corrections of both projector sets; the router weights $\pi_{a,\nu}$ mix the four environment readouts; a hypernetwork maps the element features to the specie potentials and to the specie projector strengths $D^Z_{lr}$.}
\label{sfig:network}
\end{figure}

\noindent\textbf{Training loss.}
The model is trained on the PBE band energies $E^{\rm PBE}_{n\mathbf k}$ of each structure at the $N_{\mathbf k}$ points of its band path; $E_{n\mathbf k}$ are the predicted energies, $n_{\rm v}$ is the highest occupied band of the valence electron count and $n_{\rm c}=n_{\rm v}+1$.
For a gapped structure $c$ the predicted bands are aligned at the valence-band maximum, $\tilde E_{n\mathbf k}=E_{n\mathbf k}-E_{\rm VBM}+E^{\rm PBE}_{\rm VBM}$ with $E_{\rm VBM}=\max_{\mathbf k}E_{n_{\rm v}\mathbf k}$, and the loss is
\begin{equation}
L_c=\frac{\sum_{(n,\mathbf k)\in W_c}w_n\,\big|\tilde E_{n\mathbf k}-E^{\rm PBE}_{n\mathbf k}\big|}{\sum_{(n,\mathbf k)\in W_c}w_n}
+\lambda_{\rm gap}\big|E_g-E_g^{\rm PBE}\big|
+\frac{\lambda_{\rm occ}}{n_{\rm v}N_{\mathbf k}}\sum_{n\le n_{\rm v}}\sum_{\mathbf k}\big|\tilde E_{n\mathbf k}-E^{\rm PBE}_{n\mathbf k}\big|.
\label{seq:loss_gapped}
\end{equation}
The window $W_c$ holds the states with $|E^{\rm PBE}_{n\mathbf k}-E^{\rm PBE}_{\rm mid}|\le2$\,eV, $E^{\rm PBE}_{\rm mid}$ being the midgap, together with the two band-edge bands at every $\mathbf k$, so that the edges are fitted whatever the gap; $w_n=5$ for $n=n_{\rm v},n_{\rm c}$ and 1 otherwise.
$E_g=\min_{\mathbf k}E_{n_{\rm c}\mathbf k}-\max_{\mathbf k}E_{n_{\rm v}\mathbf k}$ is the signed predicted gap, negative when the bands overlap, so that a closed gap still has a gradient; $\lambda_{\rm gap}=1$ and $\lambda_{\rm occ}=0.3$.
For gapped structures with open-shell \textit{d} atoms, whose labels include the lowest $5N_d$ conduction bands, $W_c$ is instead $[E^{\rm PBE}_{\rm VBM}-h_c,\,E^{\rm PBE}_{\rm CBM}+h_c]$ with $h_c=\max(2\,{\rm eV},\,E^{\rm PBE}_{\rm top}-E^{\rm PBE}_{\rm CBM})$, $E^{\rm PBE}_{\rm top}$ the top of the empty \textit{d} manifold.
For a metal the window $M_c$ holds the states within 2\,eV of the PBE Fermi level, the prediction is aligned by the rigid shift $s_c=\operatorname{median}_{(n,\mathbf k)\in M_c}\big(E^{\rm PBE}_{n\mathbf k}-E_{n\mathbf k}\big)$, which is not differentiated, and there is no gap term:
\begin{equation}
L_c=\frac{1}{|M_c|}\sum_{(n,\mathbf k)\in M_c}\big|E_{n\mathbf k}+s_c-E^{\rm PBE}_{n\mathbf k}\big|
+\frac{\lambda_{\rm occ}}{n_{\rm v}N_{\mathbf k}}\sum_{n\le n_{\rm v}}\sum_{\mathbf k}\big|E_{n\mathbf k}+s_c-E^{\rm PBE}_{n\mathbf k}\big|.
\label{seq:loss_metal}
\end{equation}
Each step draws a batch $\mathcal B$ of $B=16$ training structures without replacement, with probability proportional to $\omega_c/\sqrt{N_{\rm sys}(c)}$, where $N_{\rm sys}(c)$ is the number of training structures in the chemical system of $c$, so that large families do not dominate, and $\omega_c$ is the stage weight of Methods (4 for the new cells above 8 sites in the 24-site stage, 1.5 for the \textit{s}/\textit{p} cells in the final stage, 1 otherwise), and minimizes
\begin{equation}
L=\frac1B\sum_{c\in\mathcal B}L_c+\lambda_{\rm res}\sum_Z\|\mathbf r_Z\|^2+\lambda_{\rm emb}\sum_Z\|\mathbf e_Z\|^2,
\label{seq:loss}
\end{equation}
where $\mathbf r_Z$ are the per-element residuals of the specie-potential coefficients ($\lambda_{\rm res}=3\times10^{-6}$) and $\mathbf e_Z$ the per-species residuals of the encoder's species embedding ($\lambda_{\rm emb}=10^{-3}$).
In the first stage, on the \textit{s}/\textit{p} compounds, a basis-consistency term $\frac{1}{4n_{\rm v}}\sum_{n\le n_{\rm v}}\sum_{\mathbf k\in K_4}\big|\Delta E_{n\mathbf k}-\operatorname{median}\Delta E\big|$, with $\Delta E_{n\mathbf k}=E_{n\mathbf k}(20\,{\rm Ry})-E_{n\mathbf k}(g_c)$ at four $\mathbf k$ points $K_4$, was added with weight 1 for one structure per GPU and step.
Gradients of $L$ follow from the generalized Hellmann--Feynman theorem.

\FloatBarrier
\section*{2. Accuracy and band structures of held-out structures}

Table~S1 gives the accuracy of the final model in three major groups, including (i) held-out structure types, (ii) the sealed chemical systems and (iii) cells of 25--60 sites. %
By group the median gap error is \gapDten\,eV (\textit{d}$^{10}$), \gapTtwo\,eV (early \textit{d}), \gapTthree\,eV (4\textit{d}/5\textit{d}), \gapTfour\,eV (open-shell 3\textit{d}, \ngapTfour\ gapped structures) and \gapSP\,eV (\textit{s}/\textit{p}).
Restricted to compounds within 0.1\,eV/atom of the convex hull, the median gap errors are \gapstableDten, \gapstableTtwo\ and \gapstableTthree\,eV for the three larger groups.
Structure types absent from training are harder (\winProto\,eV), and the difference is concentrated in one family.
Held-out structure types containing 4\textit{d}/5\textit{d} elements are dominated by metastable mixed Mo/W chalcogenide stacks (141 of 314 structures, mostly 0.07--0.23\,eV/atom above the hull), with a median band error of 0.22\,eV; for the structures within 0.05\,eV/atom of the hull the error is 0.067\,eV.
The sealed chemical systems, set aside before training and scored once with the final model, confirm these numbers: over \nSealed\ structures the median band error is \winSealed\,eV (\leSealed\% within 0.1\,eV), between \winSealedTthree\ and \winSealedSP\,eV in every group, and the gap error \gapSealed\,eV over \ngapSealed\ gapped structures.
Larger cells of 25--60 sites and up to about 25,000 plane waves, beyond the 24 sites and 8,700 plane waves of training, are somewhat harder. For these cases, the band and gap errors are \winBigSP\ and \gapBigSP\,eV for \nBigSP\ held-out \textit{s}/\textit{p} structures and \winBigD\ and \gapBigD\,eV for \nBigD\ with \textit{d} elements, the gap error rising most for 4\textit{d}/5\textit{d} compounds (\gapBigDTthree\,eV). This suggests that the model may not have  fully learned either the atomic environment or long-range interactions of the larger cells. To further boost the model's performance, one could consider including more large-cell structures in the training set or employing data augmentation techniques to better capture the physics of these systems.

\begin{table}[htbp]
\caption{\textbf{Accuracy on held-out structures}. Band error is the mean absolute error of the band energies within $\pm2$\,eV of midgap or the Fermi level, median over structures. Gap error is the median absolute error of the band gap over gapped structures. A structure counts in every group whose elements it contains: \textit{d}$^{10}$ (Cu, Ag, Au, Zn, Cd, Hg), early \textit{d} (Sc, Y, Ti, Zr, Hf, V, Nb, Ta), 4\textit{d}/5\textit{d} (Mo, Tc, Ru, Rh, Pd, W, Re, Os, Ir, Pt), open-shell 3\textit{d} (Cr, Mn, Fe, Co, Ni). %
}
\label{stab:accuracy}
\centering
\small
\begin{tabular}{@{}lrccrc@{}}
\toprule
Set & Structures & Band error (eV) & $\le0.1$\,eV & Gapped & Gap error (eV) \\
\midrule
Held-out chemical systems & \nchemAll & \winAll & \leAll\% & \ngapAll & \gapAll \\
\quad only \textit{s}/\textit{p} elements & \nchemSP & \winSP & \leSP\% & \ngapSP & \gapSP \\
\quad with \textit{d}$^{10}$ elements & \nchemDten & \winDten & \leDten\% & \ngapDten & \gapDten \\
\quad with early \textit{d} elements & \nchemTtwo & \winTtwo & \leTtwo\% & \ngapTtwo & \gapTtwo \\
\quad with 4\textit{d}/5\textit{d} elements & \nchemTthree & \winTthree & \leTthree\% & \ngapTthree & \gapTthree \\
\quad with open-shell 3\textit{d} elements & \nchemTfour & \winTfour & \leTfour\% & \ngapTfour & \gapTfour \\
Held-out structure types & \nproto & \winProto & \leProto\% & \ngapProto & \gapProto \\
Sealed chemical systems & \nSealed & \winSealed & \leSealed\% & \ngapSealed & \gapSealed \\
\midrule
Held-out, 25--60 sites, only \textit{s}/\textit{p} & \nBigSP & \winBigSP & \leBigSP\% & \ngapBigSP & \gapBigSP \\
Held-out, 25--60 sites, with \textit{d} & \nBigD & \winBigD & \leBigD\% & \ngapBigD & \gapBigD \\
\bottomrule
\end{tabular}
\end{table}

Fig.~S2 shows band structures of held-out compounds, including the median and 90th-percentile \textit{s}/\textit{p} structures, the median structure of each \textit{d} group, and the 75th- and 90th-percentile \textit{d} structures.

\begin{figure}[htbp]
\centering
\includegraphics[width=0.99\textwidth]{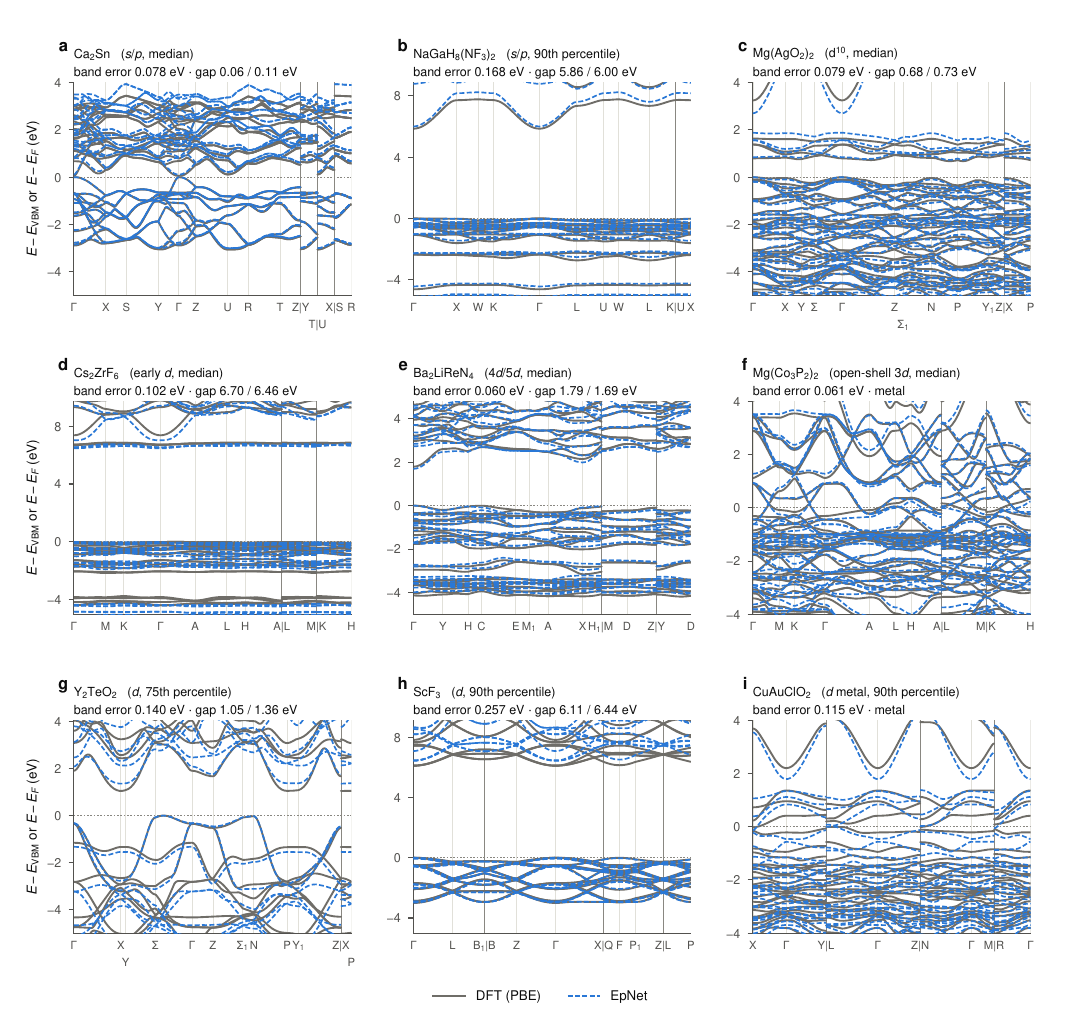}
\caption{\textbf{Band structures of held-out compounds.} PBE (gray) and EpNet (blue, dashed) for structures whose chemical systems are absent from training, chosen by rank among the held-out structures of up to 3,000 plane waves: \textbf{a,b}, the median and 90th-percentile gapped \textit{s}/\textit{p} compounds; \textbf{c,d}, the median gapped \textit{d}$^{10}$ and early-\textit{d} compounds; \textbf{e,f}, the median 4\textit{d}/5\textit{d} and open-shell 3\textit{d} structures; \textbf{g,h}, the 75th- and 90th-percentile gapped structures with \textit{d} elements; \textbf{i}, the 90th-percentile metal with \textit{d} elements. %
}
\label{sfig:gallery}
\end{figure}

\FloatBarrier
\section*{3. Atomic construction of \textit{d} channels}

We first tested whether the operator used for the \textit{s}/\textit{p} elements, with its $l=2$ Kleinman--Bylander projector placed at the \textit{d}-orbital radius, can represent \textit{d}-electron band structures at all.
Fitted to each of 45 small crystals spanning the \textit{d} block separately, it reached 0.1\,eV on at most 12 of them.
This motivated us to consider the atomic construction of \textit{d} orbitals.

Fig.~S3 shows the tests of that construction.
For seven \textit{d}$^{10}$ compounds we placed the \textit{d} element at its atomic construction, with no environment terms, next to anion potentials trained on \textit{s}/\textit{p} compounds only (Fig.~S3d,e).
A norm-conserving \textit{d} orbital, renormalized inside the core radius, scales up the all-electron tail (2.9 times for Cu, see Fig.~S3a), and the \textit{d} bands come out 3.7--10.6 times too wide even after their center is placed correctly (Fig.~S3e).
Keeping the tail and restoring the norm with the overlap operator reduces the \textit{d}--\textit{d} overlap to its physical size.
Two further ingredients are each needed: the atom's own local potential (Fig.~S3b) and a second reference energy (Fig.~S3c).
With all three, the zero-shot \textit{d}-band widths come within 1.13--1.26 times PBE for all seven compounds, with band errors of 0.05--0.35\,eV before any training.
For all 29 \textit{d} elements the constructed pseudo-atom reproduces the all-electron \textit{d} level within 0.04\,eV, with a positive-definite overlap and no spurious bound states\cite{Gonze1991}.
Table~S2 lists the parameters of the channels of every element, and equations~\eqref{seq:phi}--\eqref{seq:VdS} below show how they enter.
On 12 of these crystals, fitted one at a time, the atomic channel reached 0.1\,eV on 9, against 5 for the learned projector; trained jointly on \textit{d}$^{10}$ compounds, it transferred to held-out chemical systems with a median band error of 0.14\,eV, against 0.87\,eV for the learned projector.

\begin{figure}[htb]
\centering
\includegraphics[width=\textwidth]{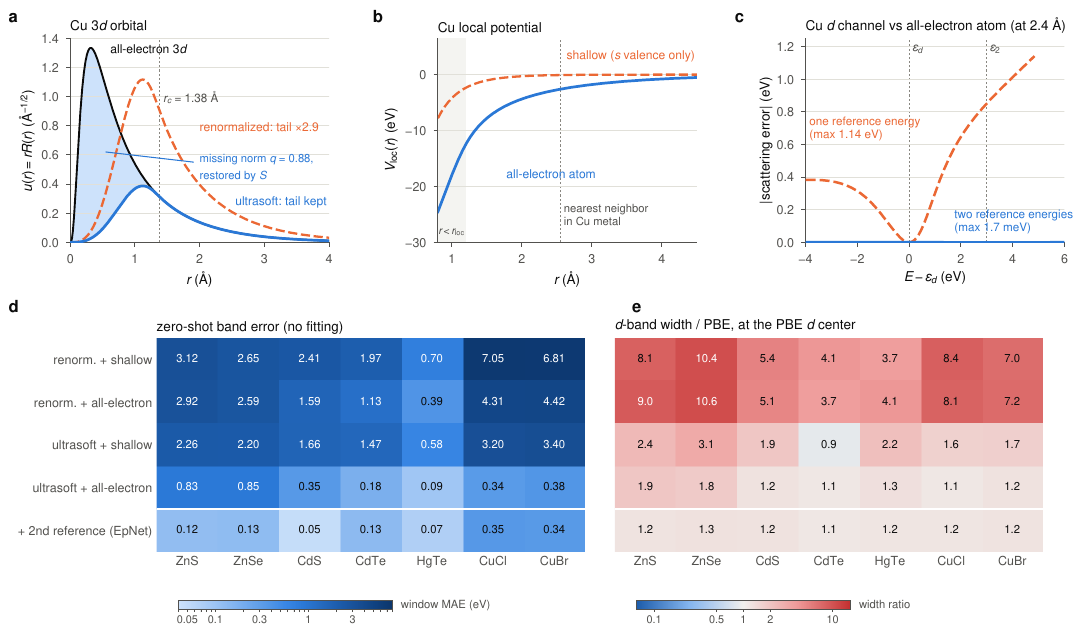}
\caption{\textbf{Building the \textit{d} channel from the atom (full construction tests).}
\textbf{a}, Cu 3\textit{d} orbital: all-electron (black), the ultrasoft pseudo-orbital that keeps the all-electron tail (blue) and a norm-conserving pseudo-orbital obtained by renormalizing it (orange). The shading is the norm restored by the overlap operator.
\textbf{b}, Cu local potential: the neutral all-electron atom beyond $r_{\rm loc}$ (blue) and the shallow \textit{s}-valence empty core (orange).
\textbf{c}, Error of the Cu \textit{d} channel in scattering energy against the all-electron atom, at 2.4\,\AA, for one and two reference energies.
\textbf{d,e}, Zero-shot test of the \textit{d} element at its atomic construction and anion potentials trained on \textit{s}/\textit{p} compounds only, for seven \textit{d}$^{10}$ compounds and five constructions. \textbf{d}, Band error in the $\pm2$\,eV window. \textbf{e}, Width of the \textit{d} band relative to PBE after shifting the \textit{d} level onto the PBE \textit{d}-band center.}
\label{sfig:dchannel}
\end{figure}

\noindent\textbf{Construction of the atomic channels.}
The parameters of Table~S2 enter the channels as follows.
The all-electron atom gives the valence levels $\varepsilon_l$ ($\varepsilon_s$, $\varepsilon_p$ and $\varepsilon_d$), which are the reference energies, and the radial orbitals $u_l(r)=rR_l(r)$, normalized to one.
Inside a core radius $r_c^l$ each orbital is replaced by a smooth, nodeless polynomial,
\begin{equation}
\phi_l(r)=\begin{cases}r^{l+1}\left(a+br^2+cr^4\right), & r<r_c^l,\\ u_l(r), & r\ge r_c^l,\end{cases}
\label{seq:phi}
\end{equation}
with $a$, $b$ and $c$ fixed by the value, slope and curvature of $u_l$ at $r_c^l$; $r_c^l$ is the smallest radius at which $\phi_l$ is nodeless and its plane-wave components above 20\,Ry carry less than 0.05\,eV of kinetic energy.
The pseudo-orbital is not renormalized: beyond $r_c^l$ it is the all-electron orbital, and inside it lacks the norm
\begin{equation}
q_l=\int_0^{r_c^l}\left(u_l^2-\phi_l^2\right)dr,
\label{seq:q}
\end{equation}
which is $q_d$ for the \textit{d} channel (0.01--0.02 for \textit{s} and \textit{p}).
The local potential is the atom's self-consistent potential $V_{\rm AE}(r)$ beyond $r_{\rm loc}$ and an even polynomial inside,
\begin{equation}
V_{\rm loc}(r)=\begin{cases}a'+b'r^2+c'r^4, & r<r_{\rm loc},\\ V_{\rm AE}(r), & r\ge r_{\rm loc},\end{cases}
\label{seq:vloc}
\end{equation}
matched in value, slope and curvature at $r_{\rm loc}$; it is the fixed table $v(q)$ in the \textit{d} element's $v_{\rm specie}$ (Methods), to which the learned part is added.
The \textit{d} channel has a second reference: the all-electron solution regular at the origin at $\varepsilon_d+\Delta\varepsilon_2$, pseudized by equation~\eqref{seq:phi} at the same $r_c^d$.
With $i,j=1,2$ labeling the two references ($\phi_1=\phi_d$, $u_1=u_d$ and $\varepsilon_1=\varepsilon_d$; $\varepsilon_2=\varepsilon_d+\Delta\varepsilon_2$) and all integrals taken within $r_{\rm cut}=\max(r_c^d,r_{\rm loc})+0.3$\,\AA,
\begin{equation}
\chi_i=\left(\varepsilon_i-T_2-V_{\rm loc}\right)\phi_i,\qquad
B_{ij}=\langle\phi_i|\chi_j\rangle,\qquad
\beta_i=\sum_j\left(B^{-1}\right)_{ji}\chi_j,
\label{seq:chi}
\end{equation}
\begin{equation}
Q_{ij}=\langle u_i|u_j\rangle-\langle\phi_i|\phi_j\rangle,\qquad
D_{ij}=B_{ij}+\varepsilon_jQ_{ij},
\label{seq:QD}
\end{equation}
where $T_l=-(\hb^2/2m)\,[d^2/dr^2-l(l+1)/r^2]$ and $Q_{11}=q_d$ (in the model $Q$ is replaced by its positive-semidefinite part, a change below 0.003).
The projectors are dual to the pseudo-orbitals, $\langle\beta_i|\phi_j\rangle=\delta_{ij}$, so the channel and its overlap,
\begin{equation}
V^d=\sum_{ij}|\beta_i\rangle D_{ij}\langle\beta_j|,\qquad
S=1+\sum_{ij}|\beta_i\rangle Q_{ij}\langle\beta_j|,
\label{seq:VdS}
\end{equation}
give $\left(T_2+V_{\rm loc}+V^d\right)\phi_i=\varepsilon_iS\phi_i$ and $\langle\phi_i|S|\phi_j\rangle=\langle u_i|u_j\rangle$.
At both reference energies, the pseudo-atom therefore has the all-electron level, the all-electron orbital beyond $r_c^d$ and the all-electron norm.
The \textit{s} and \textit{p} channels are the one-reference case, $\beta_l=\chi_l/a_l$ with $a_l=\langle\phi_l|\chi_l\rangle$, $D_l=a_l+\varepsilon_lq_l$ and $Q_l=q_l$.

In the crystal, $\beta_i(r)$ with the spherical harmonics $Y_{2m}$ is transformed to $|\mathbf k+\mathbf G|$ and placed on every atom, and equations~\eqref{seq:chi}--\eqref{seq:VdS} are re-evaluated with the current $V_{\rm loc}$ at every training step, so that they hold throughout.
Two corrections are learned on top of them: (i) a \textit{d}-level shift $\delta$ per element and per atom, applied as $D_{ij}\to D_{ij}+\delta\,\langle u_i|u_j\rangle$ so that it does not enter $B$; and (ii), for the seven elements with a value of $e_{\max}$, a stretch of the orbitals, $\phi_i(r)\to\lambda^{-1/2}\phi_i(r/\lambda)$ with $\lambda=\exp[\ln(1+e_{\max})\tanh x]$ and $x$ learned, so that $(1+e_{\max})^{-1}\le\lambda\le1+e_{\max}$. In addition, the augmentation term $D_{ij}\to D_{ij}+\Delta V_aQ_{ij}$ is added; it has no parameters of its own.
$\Delta\varepsilon_2$ and $e_{\max}$ were set for conditioning before any training: $\Delta\varepsilon_2$ is the offset among 3, 4, 5 and 6\,eV that minimizes the largest eigenvalue of $QG$, with $G_{ij}=\langle\beta_i|\beta_j\rangle$, and $e_{\max}$ the largest stretch (in steps of 0.0125, up to 0.15) over which the largest eigenvalues of $QG$ and $|D|G$ stay within three times their atomic values; below 0.025 the stretch is fixed.
The last two columns of Table~S2 are checks, not inputs: the pseudo-atom built from equations~\eqref{seq:phi}--\eqref{seq:VdS} is solved in a 20\,Ry plane-wave basis, $\delta\varepsilon_d$ is the error of its \textit{d} level against the all-electron one, and $S_{\min}$ is the smallest eigenvalue of its overlap matrix, which must be positive for $H\psi=ES\psi$ to be well posed.

\begin{table}[htbp]
\caption{\textbf{Atomic channels of the 29 \textit{d} elements} (the tables the model uses; built from the neutral,
scalar-relativistic all-electron atom in the local-density approximation, as in equations~\eqref{seq:phi}--\eqref{seq:VdS}). $\varepsilon_s$, $\varepsilon_p$,
$\varepsilon_d$: reference energies of the \textit{s}, \textit{p} and \textit{d} channels, the all-electron levels of
the atom; $r_c$: pseudization radii; $q_d$: norm deficit of the \textit{d} pseudo-orbital, restored by the overlap
operator; $r_{\rm loc}$: radius beyond which the local potential is the atom's own; $\Delta\varepsilon_2$: second
\textit{d} reference energy above $\varepsilon_d$; $e_{\max}$: bound of the learned orbital stretch $\lambda$, $(1+e_{\max})^{-1}\le\lambda\le1+e_{\max}$;
--: not learned ($\lambda=1$); $\delta\varepsilon_d$: error of the pseudo-atom's \textit{d} level at 20\,Ry against the all-electron one;
$S_{\min}$: smallest eigenvalue of the pseudo-atom's overlap.}
\label{stab:dchannels}
\centering
\footnotesize
\setlength{\tabcolsep}{4pt}
\begin{tabular}{@{}lrrrrrrrrrcrr@{}}
\toprule
 & $\varepsilon_s$ & $\varepsilon_p$ & $\varepsilon_d$ & $r_c^s$ & $r_c^p$ & $r_c^d$ & $q_d$ & $r_{\rm loc}$ & $\Delta\varepsilon_2$ & $e_{\max}$ & $\delta\varepsilon_d$ & $S_{\min}$ \\
 & (eV) & (eV) & (eV) & (\AA) & (\AA) & (\AA) & & (\AA) & (eV) & & (meV) & \\
\midrule
Sc & $-$4.30 & $-$1.52 & $-$3.43 & 1.19 & 1.23 & 1.22 & 0.48 & 1.4 & 6 & 0.088 & 28 & 0.97 \\
Ti & $-$4.60 & $-$1.54 & $-$4.46 & 1.11 & 1.14 & 1.24 & 0.59 & 1.2 & 3 & -- & 18 & 0.95 \\
V & $-$4.85 & $-$1.53 & $-$5.38 & 1.06 & 1.07 & 1.27 & 0.68 & 1.2 & 3 & -- & 15 & 0.96 \\
Cr & $-$4.20 & $-$1.08 & $-$3.13 & 1.01 & 1.03 & 1.24 & 0.65 & 1.2 & 3 & -- & 16 & 0.96 \\
Mn & $-$5.29 & $-$1.47 & $-$7.01 & 0.95 & 0.95 & 1.38 & 0.82 & 1.2 & 3 & -- & 9 & 0.98 \\
Fe & $-$5.49 & $-$1.44 & $-$7.75 & 0.90 & 0.90 & 1.38 & 0.86 & 1.2 & 3 & -- & 8 & 0.98 \\
Co & $-$5.68 & $-$1.39 & $-$8.46 & 0.87 & 0.87 & 1.38 & 0.88 & 1.2 & 3 & -- & 7 & 0.99 \\
Ni & $-$5.86 & $-$1.34 & $-$9.14 & 0.82 & 0.82 & 1.38 & 0.90 & 1.2 & 3 & -- & 6 & 0.99 \\
Cu & $-$4.86 & $-$0.79 & $-$5.33 & 0.79 & 0.82 & 1.38 & 0.88 & 1.2 & 3 & -- & 6 & 0.99 \\
Zn & $-$6.22 & $-$1.23 & $-$10.43 & 0.77 & 0.77 & 1.38 & 0.93 & 1.2 & 3 & -- & 5 & 1.00 \\
\midrule
Y & $-$4.23 & $-$1.50 & $-$2.65 & 1.36 & 1.49 & 0.77 & 0.02 & 1.8 & 3 & 0.025 & 39 & 1.00 \\
Zr & $-$4.59 & $-$1.55 & $-$3.74 & 1.30 & 1.39 & 1.16 & 0.21 & 1.8 & 3 & -- & 33 & 0.99 \\
Nb & $-$4.20 & $-$1.22 & $-$3.21 & 1.23 & 1.33 & 1.19 & 0.27 & 1.4 & 3 & 0.15 & 30 & 0.92 \\
Mo & $-$4.33 & $-$1.16 & $-$3.94 & 1.17 & 1.26 & 1.22 & 0.37 & 1.4 & 3 & 0.15 & 28 & 0.92 \\
Tc & $-$5.26 & $-$1.52 & $-$6.81 & 1.11 & 1.18 & 1.27 & 0.52 & 1.2 & 3 & -- & 22 & 0.93 \\
Ru & $-$4.52 & $-$1.04 & $-$5.40 & 1.08 & 1.15 & 1.27 & 0.55 & 1.2 & 3 & -- & 21 & 0.93 \\
Rh & $-$4.59 & $-$0.97 & $-$6.15 & 1.06 & 1.10 & 1.32 & 0.64 & 1.2 & 3 & -- & 16 & 0.94 \\
Pd & $-$3.62 & $-$0.36 & $-$4.31 & 1.03 & 1.08 & 1.30 & 0.64 & 1.2 & 3 & -- & 17 & 0.94 \\
Ag & $-$4.72 & $-$0.85 & $-$7.66 & 0.98 & 1.03 & 1.38 & 0.76 & 1.2 & 3 & -- & 13 & 0.95 \\
Cd & $-$5.96 & $-$1.33 & $-$11.90 & 0.93 & 0.98 & 1.38 & 0.83 & 1.2 & 3 & -- & 12 & 0.95 \\
\midrule
Hf & $-$5.26 & $-$1.47 & $-$2.88 & 1.18 & 1.35 & 1.14 & 0.15 & 1.6 & 3 & 0.15 & 36 & 0.98 \\
Ta & $-$5.62 & $-$1.48 & $-$3.79 & 1.16 & 1.29 & 1.19 & 0.23 & 1.4 & 3 & 0.15 & 30 & 0.92 \\
W & $-$5.91 & $-$1.47 & $-$4.69 & 1.11 & 1.26 & 1.22 & 0.31 & 1.2 & 3 & 0.15 & 28 & 0.93 \\
Re & $-$6.16 & $-$1.44 & $-$5.58 & 1.08 & 1.20 & 1.24 & 0.39 & 1.2 & 3 & -- & 26 & 0.92 \\
Os & $-$6.38 & $-$1.40 & $-$6.47 & 1.06 & 1.17 & 1.27 & 0.47 & 1.2 & 3 & -- & 23 & 0.92 \\
Ir & $-$6.59 & $-$1.35 & $-$7.36 & 1.01 & 1.12 & 1.30 & 0.54 & 1.2 & 3 & -- & 21 & 0.92 \\
Pt & $-$5.96 & $-$0.96 & $-$6.38 & 1.01 & 1.10 & 1.30 & 0.56 & 1.2 & 3 & -- & 20 & 0.93 \\
Au & $-$6.08 & $-$0.89 & $-$7.12 & 0.98 & 1.08 & 1.38 & 0.67 & 1.2 & 3 & -- & 16 & 0.93 \\
Hg & $-$7.13 & $-$1.19 & $-$10.08 & 0.93 & 1.03 & 1.38 & 0.73 & 1.2 & 3 & -- & 16 & 0.94 \\
\bottomrule
\end{tabular}
\end{table}

\FloatBarrier
\section*{4. Cost against cell size}

We selected 24 held-out structures, 4 per size range (1--2, 3--4, 5--8, 9--12, 13--16 and 17--24 sites), half of them with only \textit{s}/\textit{p} elements.
They were computed with QE 7.3 on their standardized primitive cells, using PBE, SSSP 1.3 efficiency pseudopotentials with their recommended cutoffs and Marzari--Vanderbilt smearing of 0.01\,Ry.
Each calculation is a self-consistent run on a $\Gamma$-centered grid with 0.25\,\AA$^{-1}$ spacing followed by a bands run on EpNet's path $\mathbf k$ points, with 16 MPI processes in 4 pools on one Intel Xeon Gold 6248R node.
The resulting gaps agree with the Materials Project's to a median of 0.018\,eV.
EpNet solves the same cells and $\mathbf k$ points, on one NVIDIA H100 and on 16 cores of the same node type.
On the CPU, the path is split into four contiguous parts run by 4 processes of 4 threads, as QE's pools split the $\mathbf k$ points; this is a median 1.5 times faster than one process of 16 threads.
EpNet times include setup and are taken after a first warm-up evaluation.
Fig.~S4 compares the wall time of a band structure in QE and EpNet for 24 held-out structures of 2--24 atoms with the same cells and $\mathbf k$ points.
Overall, EpNet on one GPU is 1--2 orders of magnitude faster than QE, and the speedup grows with cell size up to 24 atoms; for the 16--24-atom cells EpNet takes 4--7\,s per band structure and QE 345--984\,s.

\begin{figure}[htbp]
\centering
\includegraphics[width=0.9\textwidth]{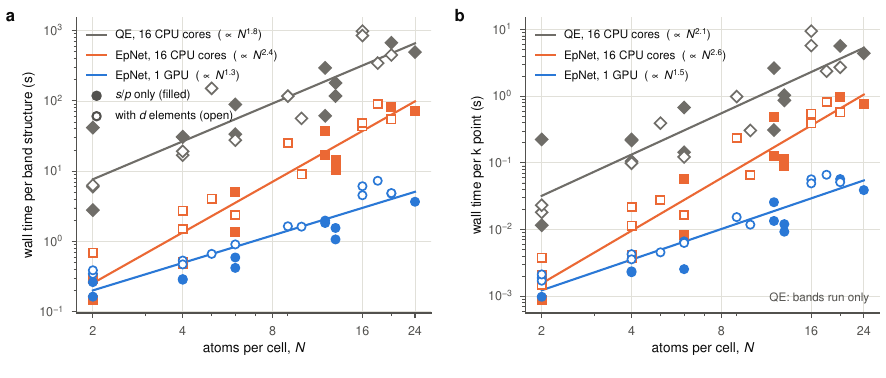}
\caption{\textbf{Band-structure cost against cell size, Quantum ESPRESSO and EpNet.} 24 held-out structures of 2--24 atoms (filled, only \textit{s}/\textit{p} elements; open, with \textit{d} elements), with the same cells and $\mathbf k$ points: QE 7.3 and EpNet on one NVIDIA H100 and on the same 16 cores. \textbf{a}, Wall time per band structure. \textbf{b}, Wall time per $\mathbf k$ point. Lines, power-law fits $t\propto N^{p}$, exponents in the legend.}
\label{sfig:qe}
\end{figure}

Fig.~S5 isolates the dependence on cell size with FCC supercells of Al and Cu of 1--512 atoms that share the same eight $\mathbf k$ points.
For the scaling test (Fig.~S5), we built $n\times n\times n$ supercells ($n=1$--8) of the Materials Project primitive cells of Al and Cu (mp-134, mp-30).
Every supercell was given the same eight Cartesian $\mathbf k$ points, $\Gamma$ and seven steps of 0.05\,\AA$^{-1}$ along a reciprocal-lattice vector of the primitive cell, and was computed with the same settings.
QE runs of Al with 512 atoms and of Cu with 216 or more atoms are omitted due to high costs. 
Time per $\mathbf k$ point is the bands-run time (QE) or the total time (EpNet) divided by eight.
Above 64 atoms, the time per $\mathbf k$ point of both codes grows as $N^{2.2}$--$N^{2.6}$, so EpNet's advantage is a constant factor rather than a lower exponent.
Overall, EpNet on the H100 is 2--3 orders of magnitude faster than QE on 16 cores for large cells.

\begin{figure}[htbp]
\centering
\includegraphics[width=0.9\textwidth]{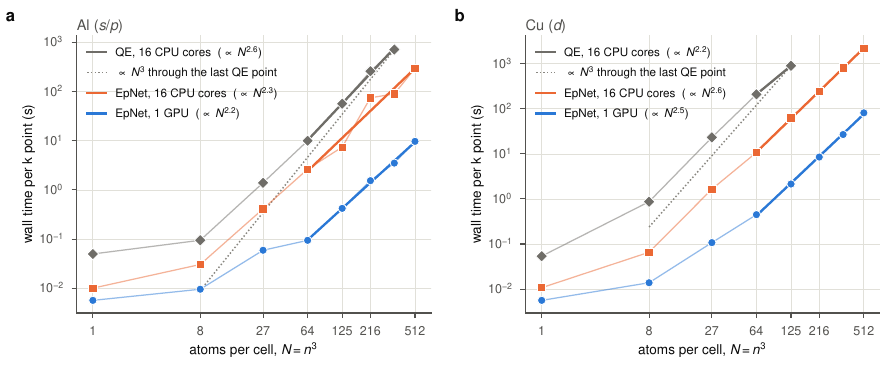}
\caption{\textbf{Cost per $\mathbf k$ point against cell size on FCC supercells.} $n\times n\times n$ supercells ($n^3=1$--512 atoms) of the primitive cells of \textbf{a}, Al and \textbf{b}, Cu, each with the same eight $\mathbf k$ points, so that only the cell size changes: QE (bands run after its self-consistent field) on 16 Intel Xeon Gold 6248R cores and EpNet on the same 16 cores and on one NVIDIA H100. QE runs of Al with 512 atoms and of Cu with 216 or more atoms are omitted, so the QE fit for Cu uses two cells (64 and 125 atoms). Lines, power-law fits over $N\ge64$; dotted, $N^3$ through the largest QE cell.}
\label{sfig:fcc}
\end{figure}

\FloatBarrier
\section*{5. Comparison with alternative models}

\begin{table}[htbp]
\caption{\textbf{Comparison of selected existing machine learning models for electronic structure calculations.} %
}
\label{stab:compare}
\centering
\footnotesize
\setlength{\tabcolsep}{3pt}
\begin{tabularx}{\textwidth}{@{}>{\raggedright\arraybackslash}p{3.4cm}>{\raggedright\arraybackslash}X>{\raggedright\arraybackslash}p{2.2cm}>{\raggedright\arraybackslash}p{2.5cm}>{\raggedright\arraybackslash}p{3.4cm}@{}}
\toprule
Approach and models & Input & Output & Training labels & Reported accuracy \\
\midrule
\textit{Property models:}\newline ALIGNN\cite{Choudhary2021ALIGNN}\newline MEGNet\cite{Chen2019,Chen2021MultiFidelity}\newline coGN\cite{Ruff2024coGN} & structure & band gap or density of states & gaps or densities of states & gap, mean error 0.16\,eV (coGN on Matbench\cite{Dunn2020Matbench}, PBE) \\
\midrule
\textit{Universal learned DFT Hamiltonians:}\newline DeepH-UMM\cite{Wang2024DeepHUMM}, UniHamGNN\cite{Zhong2024UniHamGNN}, NextHAM\cite{Yin2026NextHAM} & structure, plus the DFT code's basis and overlap matrices or a starting Hamiltonian & Hamiltonian in an atomic-orbital basis & self-consistent Hamiltonian matrices & matrix elements, 1.3--2.2\,meV \\
\midrule
\textit{Tight binding:}\newline DFTB+\cite{Hourahine2020DFTBplus} with PTBP\cite{Cui2024PTBP}, ThreeBodyTB\cite{Garrity2023TB3}\newline g-xTB\cite{Froitzheim2025gxTB} & structure & Hamiltonian in a minimal basis & DFT energies and band structures & gap, mean error 0.67--0.82\,eV against DFT\cite{Park2025CHIPSTB} \\
\midrule
\textit{Learned EPMs:}\newline Kim and Son\cite{KimSon2024}\newline Kang et~al.\cite{Kang2025MLEPM}\newline Lin et~al.\cite{Lin2025} & structure & plane-wave Hamiltonian & DFT potentials, or band energies (DFT or \textit{GW}) & single elements or material families \\
\midrule
\textbf{EpNet} (this work) & structure & plane-wave Hamiltonian & band energies & held-out chemical systems: band error \winAll\,eV, gap \gapAll\,eV \\
\bottomrule
\end{tabularx}
\end{table}

\FloatBarrier
\section*{6. Limitations of the current EpNet model}

Among the \textit{s}/\textit{p} elements the first-row elements H, F, N, O and C have the largest held-out errors (main-text Fig.~2a), and on Alexandria the gaps of structures with hydrogen are the least accurate (0.19\,eV), the worst being complex hydrides of \textit{d} metals. 

Among the \textit{d} elements, the largest gap errors are from \textit{d}$^0$ oxides and fluorides of high coordination, such as TiO$_2$ (Fig.~S6a,b). In these structures, the \textit{d} level does not drop as far as in PBE, and the gap is too large: in fluorite TiO$_2$ (0.31\,eV/atom above the hull) the O 2\textit{p} valence bands are reproduced to 0.03\,eV, but the empty Ti 3\textit{d} bands lie 0.76\,eV too high, and the gap is 1.78\,eV against 1.06\,eV.
These polymorphs lie 0.3--0.9\,eV/atom above the hull and have no counterpart in the training data.

Another limit is the layered Mo/W chalcogenide stacks. Almost all of these cells (663 of the 668 Mo/W chalcogenides of space group $P3m1$) are bilayer or trilayer slabs, separated from their periodic images by about 13\,\AA\ of vacuum.
In the held-out bilayer MoWSe$_4$ of Fig.~S6c,d, the direct gap at $\Gamma$ is 0.64\,eV too small, because the lowest conduction band there is 0.61\,eV too low, while the direct gap at K is 0.15\,eV too small and the fundamental gap is 1.05\,eV against 1.28\,eV.

The first two groups' errors are likely to be fixable by including representative structures in the training set, while the last group may require a separate treatment of van der Waals interactions for the slabs. Finally, the charges of the \textit{d} atoms are effective parameters rather than physical charges (main text, Fig.~4d). If the charges need to be physical for further applications (e.g., the electron density extraction from learned Hamiltonian), the models need to be retrained.

\begin{figure}[ht]
\centering
\includegraphics[width=\textwidth]{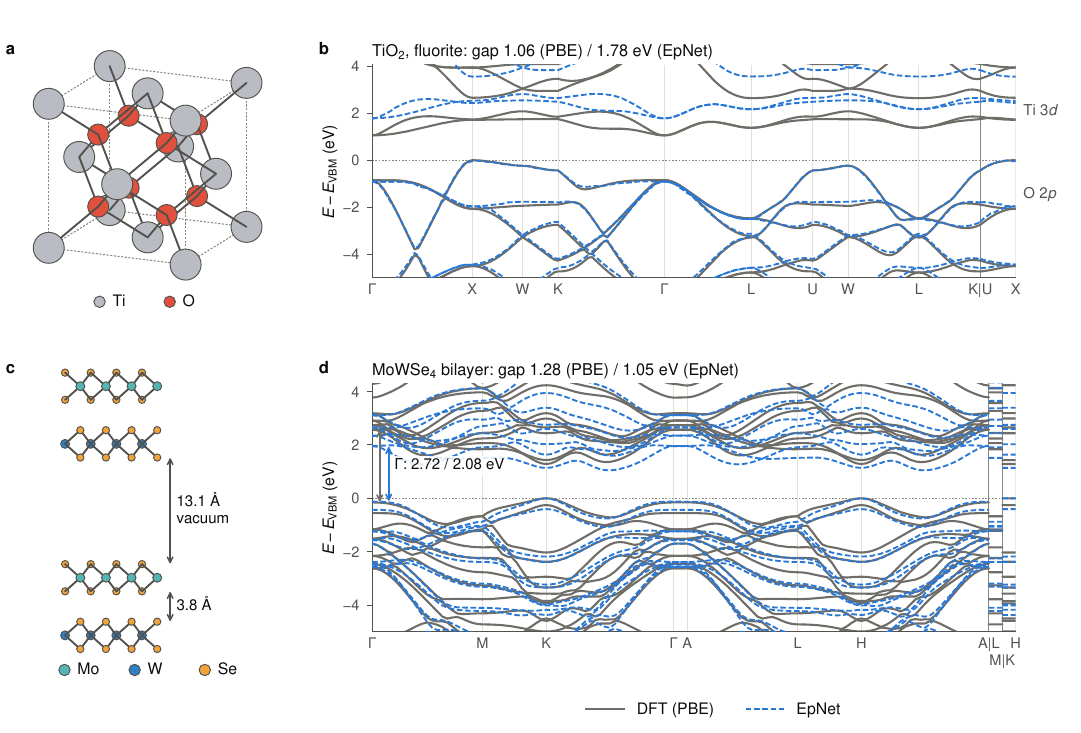}
\caption{\textbf{Two limits of the model, on held-out chemical systems.} \textbf{a}, TiO$_2$ in the fluorite structure (mp-1008677, $Fm\bar{3}m$, Ti eightfold), conventional cell. \textbf{b}, Its band structure, EpNet (blue, dashed) against PBE (gray), aligned at the valence-band maximum. \textbf{c}, MoWSe$_4$ (mp-1023928), a MoSe$_2$/WSe$_2$ bilayer, viewed edge-on to the layers over two cells along the layer normal, with the vacuum between periodic images and the van der Waals gap marked. \textbf{d}, Its band structure, as in \textbf{b}; arrows, the direct gap at $\Gamma$.}
\label{sfig:limits}
\end{figure}

\end{document}